\documentclass{article}%
\usepackage{amsmath,amsthm}%
\usepackage{amsfonts, color}%
\usepackage{amssymb}%
\usepackage{graphicx}
\usepackage{verbatim} 
\usepackage[T1]{fontenc} 
\usepackage[utf8]{inputenc} 
\usepackage{authblk} 
\usepackage{booktabs}
\newtheorem{theorem}{Theorem}

\newtheorem{corollary}{Corollary}

\newtheorem{definition}{Definition}

\newtheorem{notation}{Notation}

\newtheorem{proposition}{Proposition}

\newtheorem{assumption}{Assumption}

\usepackage[margin=1in]{geometry}

\newcommand{\bdsm}{\boldsymbol}

\begin{document}

\title{A Linear Model for Interval-valued Data}
\author[1]{Yan Sun\thanks{yan.sun@usu.edu}} 
\author[2]{Dan Ralescu\thanks{dan.ralescu@uc.edu}} 
\affil[1]{Department of Mathematics $\&$ Statistics\\  
Utah State University\\
3900 Old Main Hill\\
Logan, Utah 84322} 
\affil[2]{Department of Mathematical Sciences\\  
University of Cincinnati\\
2600 Clifton Avenue\\
Cincinnati, Ohio 45220
}

\date{}
\maketitle

\begin{abstract}
Interval-valued linear regression has been investigated for some time. One of the critical issues is optimizing the balance between model flexibility and interpretability. This paper proposes a linear model for interval-valued data based on the affine operators in the cone $\mathcal{C} = \{ (x, y) \in \mathbb{R}^2 | x \leq y\}$. The resulting new model is shown to have improved flexibility over typical models in the literature, while maintaining a good interpretability. The least squares (LS) estimators of the model parameters are provided in a simple explicit form, which possesses a series of nice properties. Further investigations into the LS estimators shed light on the positive restrictions of a subset of the parameters and their implications on the model validity. A simulation study is presented that supports the theoretical findings. An application to a real data set is also provided to demonstrate the applicability of our model.
\end{abstract}


\section{Introduction}\label{intro}
Recently there has been an increasing interest in the linear regression for interval-valued data. See Diamond(1990), K\"orner and N\"ather (1998) , Gil et al. (2002, 2007), Manski and Tamer (2002), Carvalho et al. (2004), Billard (2007), Gonz\'alez-Rodr\'iguez et al. (2007), Lima Neto and De Carvalho (2008, 2010), Blanco-Fern\'andez et al. (2011), Cattaneo and Wiencierz (2012), for a partial list of references. Existing models have been developed mainly in two directions. In the first direction, separate point-valued linear regression models are fitted to the center and range (or the lower and upper bounds), respectively, treating the intervals essentially as bivariate vectors. Examples belonging to this category include the center method by Billard and Diday (2000), the MinMax method by Billard and Diday (2002), the (constrained) center and range method by Lima Neto and De Carvalho (2008, 2010), and the model M by Blanco-Fern\'andez et al. (2011). The second direction is to view the intervals as subsets in $\mathbb{R}$ and study their linear relationship in the framework of random sets. Investigations along this direction include Diamond (1990), Gil et al. (2001, 2002), Gil et al. (2007), Gonz\'alez-Rodr\'iguez (2007), and Sun and Li (2014), among others. In this paper, we propose a new linear model for interval-valued data that aims at connecting the two directions and achieving improved flexibility.  

To facilitate our presentation, let us give a brief introduction on the theoretical framework of random sets. 
Let $(\Omega,\mathcal{L},P)$ be a probability space. Denote by $\mathcal{K}\left(\mathbb{R}^d\right)$ or $\mathcal{K}$ the collection of all non-empty compact subsets of $\mathbb{R}^d$. In the space $\mathcal{K}$, a linear structure is defined by Minkowski addition and scalar multiplication, i.e.,
\begin{equation}\label{def:int-linear}
  A+B=\left\{a+b: a\in A, b\in B\right\},\ \ \ \ \lambda A=\left\{\lambda a: a\in A\right\},
\end{equation}
$\forall A, B\in\mathcal{K}$ and $\lambda\in\mathbb{R}$. 
A random compact set is a Borel measurable function $A: \Omega\rightarrow\mathcal{K}$, $\mathcal{K}$ being equipped with the Borel $\sigma$-algebra induced by the Hausdorff metric. For each $X\in\mathcal{K}\left(\mathbb{R}^d\right)$, the function defined on the unit sphere $S^{d-1}$:
\begin{equation*}
  s_X\left(u\right)=\sup_{x\in X}\left<u, x\right>,\ \ \forall u\in S^{d-1}
\end{equation*}
is called the support function of X. If $A(\omega)$ is convex almost surely, then $A$ is called a random compact convex set. (See Molchanov 2005, p.21, p.102.) The collection of all compact convex subsets of $\mathbb{R}^d$ is denoted by $\mathcal{K}_{\mathcal{C}}\left(\mathbb{R}^d\right)$ or $\mathcal{K}_{\mathcal{C}}$. Especially, when $d=1$, $\mathcal{K}_{\mathcal{C}}(\mathbb{R})$ contains all the non-empty bounded closed intervals in $\mathbb{R}$. A measurable function $X: \Omega\rightarrow\mathcal{K}_\mathcal{C}\left(\mathbb{R}\right)$ is called a random interval. Much of the random sets theory has focused on compact convex sets (see, e.g., Artstein and Vitale (1975), Aumann (1965), and Lyashenko (1982, 1983)). Let $\mathcal{S}$ be the space of support functions of all non-empty compact convex subsets in $\mathcal{K}_{\mathcal{C}}$. Then, $\mathcal{S}$ is a Banach space equipped with the $L_2$ metric 
\begin{equation*}
  \|s_X(u)\|_2=\left[d\int_{S^{d-1}}|s_X(u)|^2\mu\left(\mathrm{d}u\right)\right]^{\frac{1}{2}},
\end{equation*}
where $\mu$ is the normalized Lebesgue measure on $S^{d-1}$. According to various embedding theorems (see R\aa dstr\"om 1952; H\"ormander 1954), $\mathcal{K}_{\mathcal{C}}$ can be embedded isometrically into the Banach space $C(S)$ of continuous functions on $S^{d-1}$, and $\mathcal{S}$ is the image of $\mathcal{K}_\mathcal{C}$ into $C(S)$. Therefore, $\delta\left(X, Y\right):=\|s_X-s_Y\|_2$, $\forall X, Y\in\mathcal{K}_\mathcal{C}$, defines an $L_2$ metric on $\mathcal{K}_\mathcal{C}$.

The central idea to constructing linear models in the random sets framework is to minimize the distance $\delta\left(Y, E(Y|X)\right)$ on the data, where $X, Y$ are random intervals and $E(Y|X)$ is a linear function of $X$ in the sense of (\ref{def:int-linear}). Such models have very nice mathematical interpretations, but the restriction to the space $\mathcal{K}_\mathcal{C}\left(\mathbb{R}\right)$ unfortunately results in a reduced flexibility from practical point of view. Notice that 
\begin{eqnarray*}
  \left(aX+b\right)^C&=&aX^C+b,\\
  \left(aX+b\right)^R&=&|a|X^R. 
\end{eqnarray*}
This implies that the slope parameters of the corresponding linear model for the center (C) and range (R) must be the same in absolute value (see, e.g., Gil et al. (2002) and Sun and Li (2014)). Such a restriction is usually relaxed in the models developed for the center and range separately. Those models typically treat an interval as a vector in the Euclidean space $\mathbb{R}^2$ and minimize the Euclidean distance $\left\|Y-E\left(Y|X\right)\right\|$ on the data. However, this approach is slightly problematic in that once the intervals are represented by vectors in $\mathbb{R}^2$, they should be modeled as such, as opposed to being broken down to the centers and ranges separately. Particularly, a linear model in $\mathbb{R}^2$ in general takes on the form 
\begin{equation}\label{model-origin}
  Y=AX+\textbf{b}+\bdsm{\epsilon},
\end{equation}
where $A$ is a $2\times 2$ coefficient matrix,  $\textbf{b}$ is a $2\times 1$ intercept vector, and $\bdsm{\epsilon}$ is a $2\times 1$ error vector. There is no reason to separate the two coordinates of $X$ by forcing $A$ to be diagonal. This problem makes the bivariate types of models hard to interpret both in $\mathcal{K}_\mathcal{C}\left(\mathbb{R}\right)$ and in $\mathbb{R}^2$. 

Our main contribution in this paper is to generalize the bivariate types of models from the literature (i.e., models in the first research direction by the preceding discussion) to the form (\ref{model-origin}), which is accomplished by embedding the space $\mathcal{K}_\mathcal{C}\left(\mathbb{R}\right)$ into $\mathbb{R}^2$, and more precisely, into the cone $\mathcal{C} = \{ (x, y) \in \mathbb{R}^2 | x \leq y\}$. As such, our proposed new linear model has generally improved flexibility over the existing models in both directions. It is also well interpretable in $\mathcal{C}$ due to the embedding. We extend the univariate model to the multiple case and derive the matrix form of the general multivariate model. The least squares (LS) estimates for the model parameters are provided in matrix form, from which a series of properties are derived. Furthermore, we give explicit analytical LS solutions for the positive parameters, which shed light on the behaviors of the LS estimators in connection with the positive restriction and the model validity. Simulation studies are carried out that produce consistent results with our theoretical findings. Finally, an application to a real data set is presented to demonstrate the applicability of our model. 

The rest of the paper is organized as follows. Section 2 formally introduces our model and discusses the associated model properties. The LS estimators and their properties are presented in Section 3, followed by a rigorous discussion on the estimation of the positive parameters in Section 4. Simulation studies are reported in Section 5, and the real data application is presented in Section 6. We give concluding remarks in Section 7. Technical proofs are collected in the Appendix. 


\section{The linear model}
\subsection{The affine operator in $\mathcal{C}$ and the univariate model in $\mathcal{K}_\mathcal{C}(\mathbb{R})$}
Assume observing an i.i.d. random sample of paired intervals $X_i=\left[X_i^{L}, X_i^{U}\right]$, $Y_i=\left[Y_i^{L}, Y_i^{U}\right]$, $i=1,\cdots,n$, where $X_i^{L}$, $Y_i^{L}$ and $X_i^{U}$, $Y_i^{U}$ are the lower and upper bounds of $X_i$ and $Y_i$, respectively. Alternatively, the interval $X_i$ can also be represented by its center $X_i^C$ and range $X_i^R$ as
\begin{eqnarray*}
  X_i^C &=& \left(X_i^U+X_i^L\right)/2,\\
  X_i^R &=& X_i^U-X_i^L,
\end{eqnarray*}
and similarly for $Y_i$. The $\delta$-metric in the space $\mathcal{K}_\mathcal{C}(\mathbb{R})$ is given by 
\begin{equation*}
  \delta\left(X_1,X_2\right)=\sqrt{\frac{1}{2}\left(X_1^L-X_2^L\right)^2+\frac{1}{2}\left(X_1^U-X_2^U\right)}.
\end{equation*}
This suggests that the metric space $\left(\mathcal{K}_\mathcal{C}(\mathbb{R}),\delta\right)$ can be embedded isometrically into the cone $\mathcal{C} = \{ (x, y) \in \mathbb{R}^2 | x \leq y\}$ equipped with the Euclidean metric. 
Therefore, we consider each interval $X=\left[X^L,X^U\right]\in\mathcal{K}_\mathcal{C}(\mathbb{R})$ to be represented by the point $\left(X^L,X^U\right)\in\mathcal{C}$. 

From the preceding discussion of embedding, we propose to construct a linear model in $\mathcal{K}_\mathcal{C}(\mathbb{R})$ based on the affine operator in $\mathcal{C}$, i.e., affine operator $T: \mathbb{R}^2 \to \mathbb{R}^2$ satisfying $T(\mathcal{C}) \subseteq \mathcal{C}$. Obviously, such affine operators are represented by
\begin{equation*}
 T\left(\begin{bmatrix}x\\ y\end{bmatrix}\right)
 =\begin{bmatrix}\alpha & \beta\\ \alpha-\gamma & \beta+\gamma\end{bmatrix}\begin{bmatrix}x\\ y\end{bmatrix}+\begin{bmatrix}\eta\\ \eta+\theta\end{bmatrix},
\end{equation*}
with $\alpha, \beta, \eta \in\mathbb{R}$ and $\gamma, \theta\geq 0$. This leads us to propose the following univariate linear model 
\begin{eqnarray}
  Y_i^L&=&\alpha X_i^L+\beta X_i^U+\eta+\epsilon_i^L,\label{uni-mod-1}\\
  Y_i^U&=&\left(\alpha-\gamma\right)X_i^L+\left(\beta+\gamma\right)X_i^U+\eta+\theta+\epsilon_i^U, \label{uni-mod-2}
\end{eqnarray}
where $\alpha, \beta, \eta \in\mathbb{R}, \gamma, \theta\geq 0$ are coefficients, and $\left\{\epsilon_i^L, \epsilon_i^U\right\}$ are i.i.d. zero mean random variables with variance $\sigma^2>0$, $i=1,\cdots, n$. 

\subsection{Collinearity preservation}
\begin{figure}[ht]
\centering
\includegraphics[height=2.000in, width=2.200in]{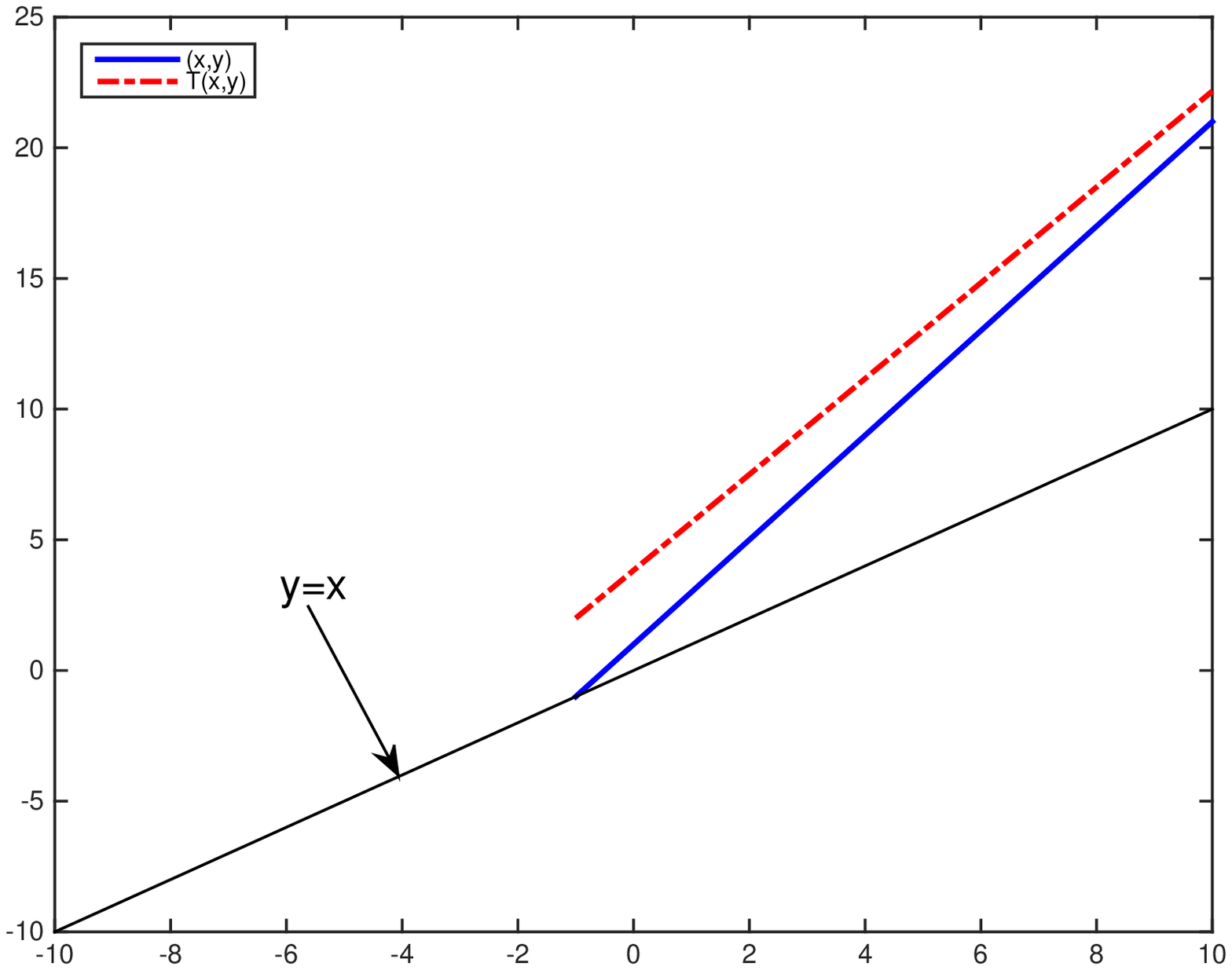}
\includegraphics[height=2.000in, width=2.200in]{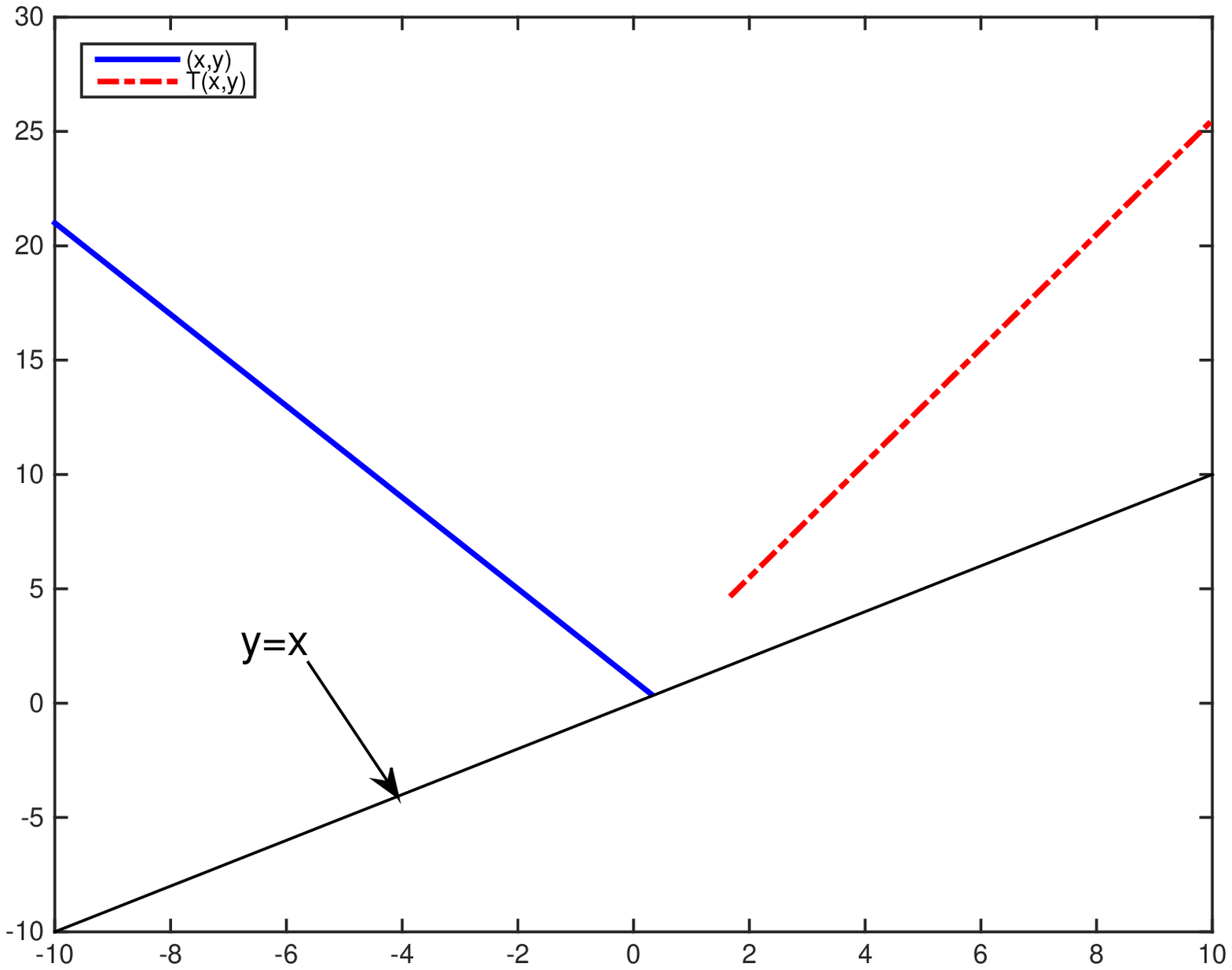}
\caption{A graphical illustration of the affine transformation in $\mathcal{C}$, which is above the line $y=x$ in $\mathbb{R}^2$. The solid line is a ray $y=ax+b,y\geq x$ in $\mathcal{C}$, and the dash-dotted line is its image by $T$ with the parameters $\alpha=-2,\beta=4,\gamma=5,\eta=1,\theta=3$. Left: $a=2,b=1$; Right: $a=-2,b=1$.}
\label{fig:affine}
\end{figure}

\begin{figure}[ht]
\centering
\includegraphics[height=1.800in, width=2.000in]{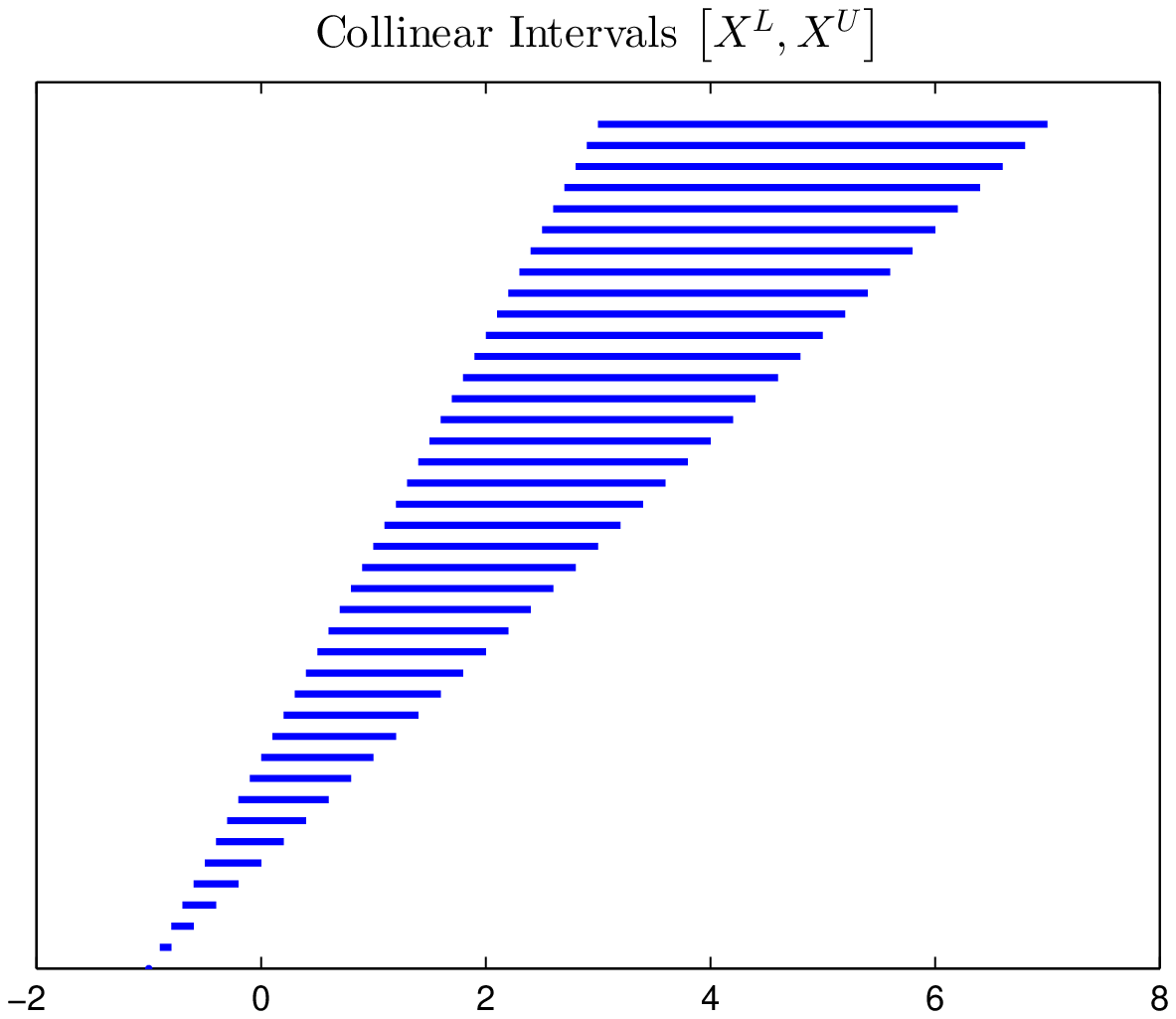}
\includegraphics[height=1.800in, width=2.000in]{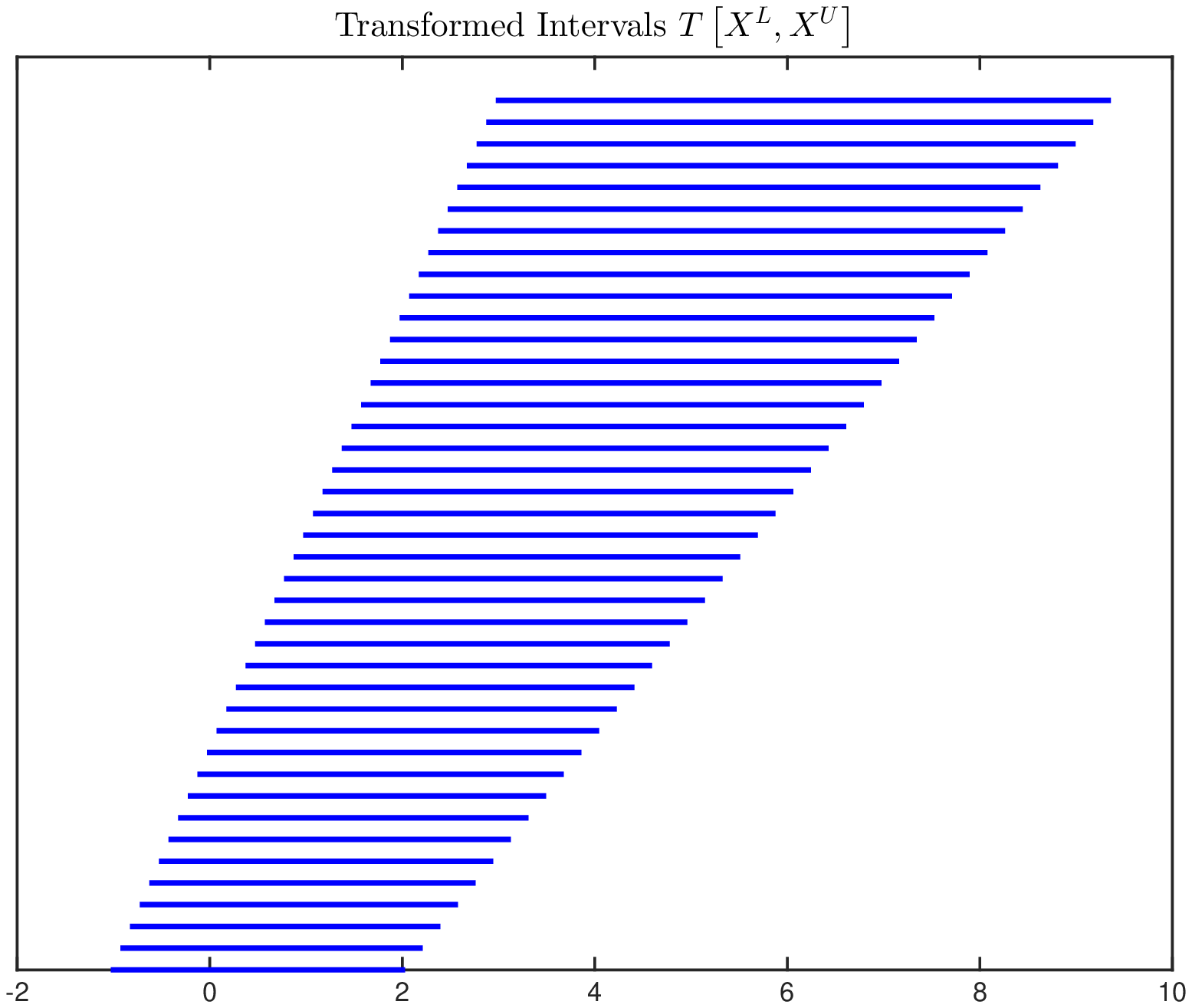}\\
\includegraphics[height=1.800in, width=2.000in]{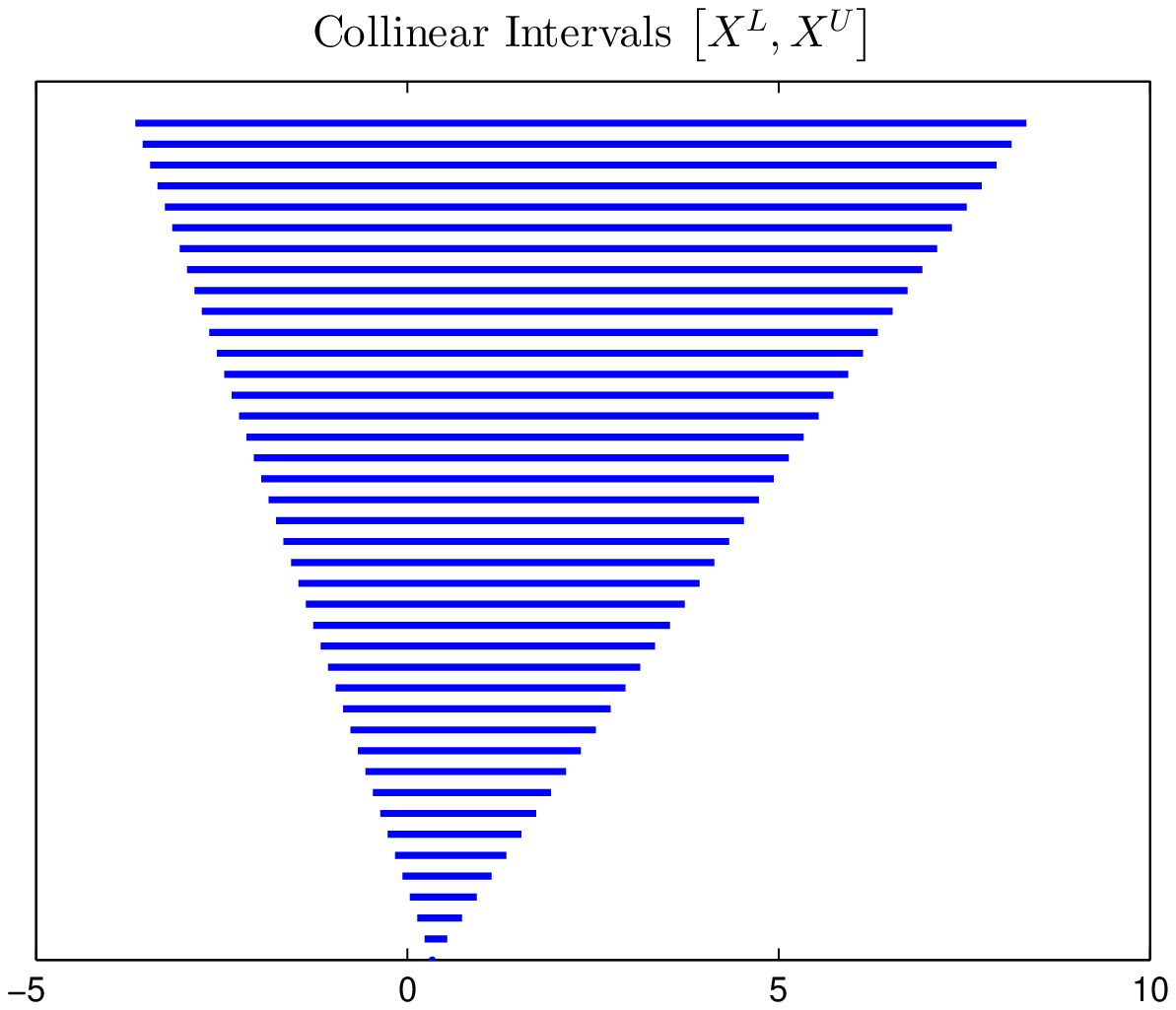}
\includegraphics[height=1.800in, width=2.000in]{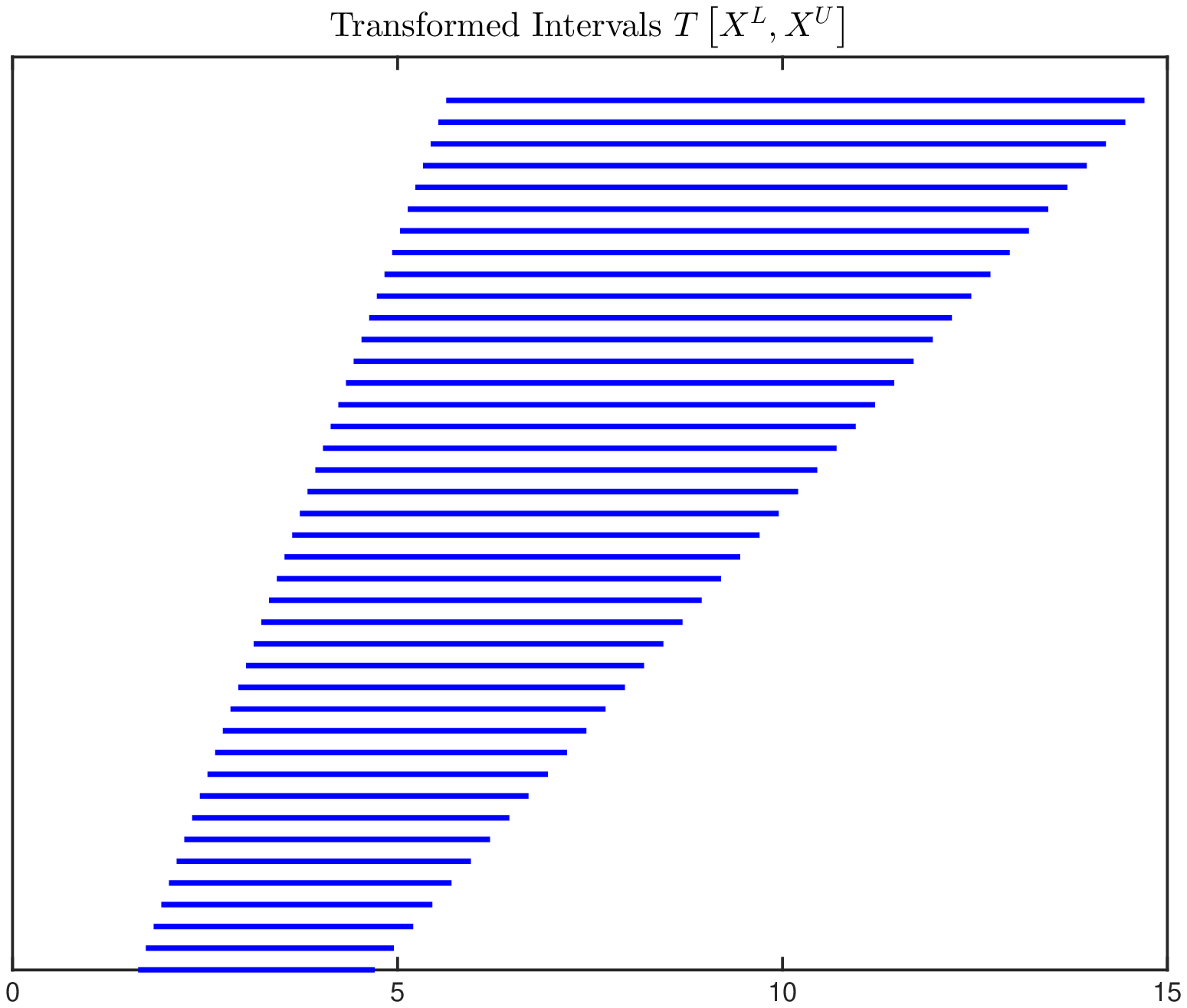}\\
\caption{Visualization of collinear intervals. Each interval is displayed as a horizontal line segments, and the intervals are elevated proportionally in order to be displayed in one plot. The left two plots are collinear intervals that satisfy equation (\ref{def-collinear-1}) with (top) $a=2,b=1$ and (bottom) $a=-2,b=1$. The right two plots are their corresponding images by $T$ with parameters $\alpha=-2,\beta=4,\gamma=5,\eta=1,\theta=3$.}
\label{fig:collinearity}
\end{figure}

The most important property of affine transformation is that it preserves collinearlity. This in the cone $\mathcal{C}$ means that points lying on a ray are still on a ray after transformation. Precisely, the operator $T$ maps the ray
\begin{equation*}
  y=ax+b,\ y\geq x
\end{equation*}
into another ray
\begin{equation*}
  T\left(y\right)=\left[1+\frac{\gamma\left(a-1\right)}{\alpha+\beta a}\right]T\left(x\right)
	+\gamma b+\theta-\frac{\gamma\left(a-1\right)}{\alpha+\beta a}\left(\beta b+\eta\right),\ 
	T\left(y\right)\geq T\left(x\right).
\end{equation*}
Figure \ref{fig:affine} gives an illustration of this effect. Considering the equivalence of the point $\left(X^L,X^U\right)\in\mathcal{C}$ and the interval $[X^L,X^U]\in\mathcal{K}_\mathcal{C}(\mathbb{R})$, we define a collection of intervals $\left\{[X_i^L,X_i^U]:i\in I\right\}$ to be collinear if their representations $\left\{\left(X_i^L,X_i^U\right):i\in I\right\}$ in $\mathcal{C}$ are on a ray. 
\begin{definition}
A collection of intervals $\left\{[X_i^L,X_i^U]:i\in I\right\}$ are said to be collinear if they satisfy the equation
\begin{equation}\label{def-collinear-1}
  X_i^U=aX_i^L+b,
\end{equation}
where $X_i^L\geq -\frac{b}{a-1}$ if $a>1$, $X_i^U\leq -\frac{b}{a-1}$ if $a<1$, and $b\geq 0$ if $a=1$. 
\end{definition}
It is easily seen that equation (\ref{def-collinear-1}) can be equivalently expressed as
\begin{equation*}
  X^R=\frac{2}{a+1}\left[\left(a-1\right)X^C+b\right].
\end{equation*}
So, we can also define collinearity in terms of the center and range of the interval.
\begin{definition}
The collinearity of a collection of intervals $\left\{[X_i^L,X_i^U]:i\in I\right\}$ is equivalently defined by 
\begin{equation}\label{def-collinear-2}
  X_i^R=cX_i^C+d,
\end{equation}
where $X_i^C\geq -\frac{d}{c}$ if $c>0$, $X_i^C\leq \frac{d}{c}$ if $c<0$, and $d\geq 0$ if $c=0$. 
\end{definition}
From these two definitions, collinearity of intervals essentially means that the upper bound changes linearly with the lower bound, or equivalently, the range changes linearly with the center. For example, it is a common situation in practice that a larger center is associated with a wider range. When this relationship is linear, the corresponding intervals are considered collinear, and such a characteristic gets preserved under the operator $T$. Figure \ref{fig:collinearity} provides a visualization of this property. In terms of modeling, if an interval-valued data $x_i=\left[x_i^{L}, x_i^{U}\right]$, $y_i=\left[y_i^{L}, y_i^{U}\right]$, $i=1,\cdots,n$, follows our model (\ref{uni-mod-1})-(\ref{uni-mod-2}), then for $x_i$'s that are collinear, their associated $y_i$'s are also collinear.  

\subsection{Comparison to other models}
\label{sec:model-compare}
As we mentioned in the introduction, our model has systematically improved flexibility over typical models in the literature. In this section, we compare our univariate model (\ref{uni-mod-1})-(\ref{uni-mod-2}) to two popular models to gain more insight into this. Consider the M model proposed by Blanco-Fern\'andez et al. (2011), and the constrained center and range method (CCRM) by Lima Neto and De Carvalho (2010). The M model is specified as
\begin{eqnarray}
  Y_i^C&=&\alpha X_i^C+\gamma+\text{mid}\epsilon_i,\label{M-1}\\
	Y_i^R&=&|\beta| X_i^R+\text{spr}\epsilon_i,\label{M-2}
\end{eqnarray}
where $\text{mid}\epsilon_i$ and $\text{spr}\epsilon_i$ are center and range of the interval-valued random error $\epsilon_i$, respectively. $\text{mid}\epsilon_i$ is assume to be a centered random variable and $\text{spr}\epsilon_i$ is assumed to be a positive random variable. On the other hand, the model of CCRM is defined as 
\begin{eqnarray}
  Y_i^C&=&\beta_0^C+\beta_1^C X_i^C+\epsilon_i^C,\label{ccrm-1}\\
	Y_i^R&=&\beta_0^R+\beta_1^R X_i^R+\epsilon_i^R,\label{ccrm-2}
\end{eqnarray}
where $\epsilon_i^C$ and $\epsilon_i^R$ are both centered random variables without any geometric interpretations. The two coefficients $\beta_0^R$, $\beta_1^R$ in the range regression equation are both restricted to be positive to ensure the positiveness of $Y_i^R$. It is easy to see that these two models are essentially equivalent with $\epsilon_i^C=\text{mid}\epsilon_i$, $\beta_o^R=\text{E}\left(\text{spr}\epsilon_i\right)$, and $\epsilon_i^R=\text{spr}\epsilon_i-\text{E}\left(\text{spr}\epsilon_i\right)$. Rewriting equations (\ref{ccrm-1})-(\ref{ccrm-2}) in terms of the lower and upper bounds, the model of CCRM is equivalently represented as
\begin{eqnarray}
  Y_i^L&=&\beta_0^C-\frac{1}{2}\beta_0^R+\frac{1}{2}\left(\beta_1^C+\beta_1^R\right)X_i^L
	+\frac{1}{2}\left(\beta_1^C-\beta_1^R\right)X_i^U+\epsilon_i^C-\frac{1}{2}\epsilon_i^R,\label{ccrm-L}\\
  Y_i^U&=&\beta_0^C+\frac{1}{2}\beta_0^R+\frac{1}{2}\left(\beta_1^C-\beta_1^R\right)X_i^L
	+\frac{1}{2}\left(\beta_1^C+\beta_1^R\right)X_i^U+\epsilon_i^C+\frac{1}{2}\epsilon_i^R,\label{ccrm-U}
\end{eqnarray}
where $\beta_0^R>0$, $\beta_1^R>0$. This compared to our model (\ref{uni-mod-1})-(\ref{uni-mod-2}) is a reduced form with the restrictions $\alpha=\beta+\gamma$. So our model has one extra degree of freedom, which will drastically expand the model flexibility. The CCRM is extended to the multiple case, from which the advantage of our general model introduced in the following gets multiplied. We will elaborate more on this in the simulation and real data application sections. 

\subsection{Matrix form of the general model}
Consider the general case involving the outcome interval $Y_i=\left[Y_i^L, Y_i^U\right]$ and $p$ interval-valued predictors $X_{j,i}=\left[X_{j,i}^L, X_{j,i}^U\right]$, $i=1,\cdots,n$; $j=1,\cdots,p$. To model $Y_i$ by a linear transformation of $\left\{X_{j,i}: j=1,\cdots,p\right\}$, we extend the univariate model (\ref{uni-mod-1})-(\ref{uni-mod-2}) to the following form:
\begin{eqnarray}
  Y_i^L&=&\sum_{j=1}^{p}\left(\alpha_jX_{j,i}^L+\beta_jX_{j,i}^U\right)+\eta+\epsilon_i^L,\label{gen-mod-1}\\
  Y_i^U&=&\sum_{j=1}^{p}\left[\left(\alpha_j-\gamma_j\right)X_{j,i}^L+\left(\beta_j+\gamma_j\right)X_{j,i}^U\right]+\eta+\theta+\epsilon_i^U,\label{gen-mod-2}
\end{eqnarray}
with $\gamma_j\geq 0$, $\text{E}\left(\epsilon_i^U\right)=\text{E}\left(\epsilon_i^L\right)=0$, and $\text{Var}\left(\epsilon_i^U\right)=\text{Var}\left(\epsilon_i^L\right)=\sigma^2>0$, $\theta\geq 0$, for $i=1,\cdots,n$ and $j=1,\cdots,p$. 

Define 
\begin{equation*}
  {\bf Y}^L=\begin{bmatrix}Y_1^L\\ Y_2^L\\ \cdot\\ \cdot\\ \cdot\\ Y_n^L\end{bmatrix},\ \ \ 
  {\bf X}_1=\begin{bmatrix}1 & X_{1,1}^L & X_{1,1}^U & X_{2,1}^L & X_{2,1}^U & \cdots & X_{p,1}^L & X_{p,2}^U\\
                                1 & X_{1,2}^L & X_{1,2}^U & X_{2,2}^L & X_{2,2}^U & \cdots & X_{p,2}^L & X_{p,2}^U\\
                                 \cdot & \cdot & \cdot & \cdot & \cdot & \cdots & \cdot & \cdot\\
                                  \cdot & \cdot & \cdot & \cdot & \cdot & \cdots & \cdot & \cdot\\
                                   \cdot & \cdot & \cdot & \cdot & \cdot & \cdots & \cdot & \cdot\\
                                1 & X_{1,n}^L & X_{1,n}^U & X_{2,n}^L & X_{2,n}^U & \cdots & X_{p,n}^L & X_{p,n}^U\end{bmatrix},
\end{equation*}
and
\begin{equation*}
   \boldsymbol{\beta}_1=\begin{bmatrix} \eta\\ \alpha_1\\ \beta_1\\
   \cdot \\ \cdot \\ \cdot \\ \alpha_p\\ \beta_p\end{bmatrix},\ \ \ 
  \boldsymbol{\epsilon}^L=\begin{bmatrix} \epsilon_1^L\\ \epsilon_2^L\\ \cdot \\ \cdot \\ \cdot \\ \cdot \\ \cdot \\
                                               \epsilon_n^L\end{bmatrix}.
\end{equation*}
Then equation (\ref{gen-mod-1}) is expressed as 
\begin{equation*}
  {\bf Y}^L={\bf X}_1\bdsm{\beta}_1+\bdsm{\epsilon}^L.
\end{equation*}
Define 
\begin{equation*}
  {\bf Y}^U=\begin{bmatrix}Y_1^U\\ Y_2^U\\ \cdot\\ \cdot\\ \cdot\\ Y_n^U\end{bmatrix},\ \ \ 
  {\bf X}_2=\begin{bmatrix}1 & X_{1,1}^R & X_{2,1}^R & \cdots & X_{p,1}^R\\
                                1 & X_{1,2}^R & X_{2,2}^R & \cdots & X_{p,2}^R\\
                                 \cdot & \cdot & \cdot & \cdot & \cdot\\
                                 \cdot & \cdot & \cdot & \cdot & \cdot\\
                                 \cdot & \cdot & \cdot & \cdot & \cdot\\
                                1 & X_{1,n}^R & X_{2,n}^R & \cdots & X_{p,n}^R\end{bmatrix},
\end{equation*}
and
\begin{equation*}
   \boldsymbol{\beta}_2=\begin{bmatrix} \theta\\ \gamma_1\\ \gamma_2\\
   \cdot \\ \cdot \\ \cdot \\ \gamma_p\end{bmatrix},\ \ \ 
  \boldsymbol{\epsilon}^U=\begin{bmatrix} \epsilon_1^U\\ \epsilon_2^U\\ \cdot \\ \cdot \\ \cdot \\ \cdot \\ 
                                               \epsilon_n^U\end{bmatrix}.
\end{equation*}
Then equation (\ref{gen-mod-2}) is rewritten as
\begin{equation*}
  {\bf Y}^U=\begin{bmatrix}{\bf X}_1 & {\bf X}_2\end{bmatrix}
  \begin{bmatrix}\bdsm{\beta}_1\\ \bdsm{\beta}_2\end{bmatrix}
    +\bdsm{\epsilon}^U.
\end{equation*}
To jointly express the model, define
\begin{equation*}
  {\bf Y}=\begin{bmatrix}{\bf Y}^L\\ {\bf Y}^U \end{bmatrix},\ \ \
  {\bf X}=\begin{bmatrix}{\bf X}_1 & {\bf 0}\\ {\bf X}_1 & {\bf X}_2 \end{bmatrix},\ \ \
  \bdsm{\beta}=\begin{bmatrix}\bdsm{\beta}_1\\ \bdsm{\beta}_2 \end{bmatrix},\ \ \ \text{and}\ \ \ 
  \bdsm{\epsilon}=\begin{bmatrix}\bdsm{\epsilon}^L\\ \bdsm{\epsilon}^U \end{bmatrix} . 
\end{equation*}
Then, the general model (\ref{gen-mod-1})-(\ref{gen-mod-2}), can be written in the matrix form
\begin{equation}\label{mod-matrix}
  \textbf{Y}=\textbf{X}\boldsymbol{\beta}+\boldsymbol{\epsilon}.
\end{equation}

\section{Least squares estimation}\label{sec:para-esti}
We define our least squares estimator $\hat{\bdsm{\beta}}$ of $\bdsm{\beta}$ as the minimizer of the sum of squared lower and upper bound errors. Namely,
\begin{equation}\label{beta-lse}
  \hat{\bdsm{\beta}}=\arg\min\left\{\sum_{i=1}^{n}\left[\left(Y_i^L-\hat{Y_i^L}\right)^2+\left(Y_i^U-\hat{Y_i^U}\right)^2\right]\right\},
\end{equation}
where
\begin{eqnarray}
  \hat{Y_i^L}&=&\sum_{j=1}^{p}\left(\alpha_jX_{j,i}^L+\beta_jX_{j,i}^U\right)+\eta,\label{pred_L}\\
  \hat{Y_i^U}&=&\sum_{j=1}^{p}\left[\left(\alpha_j-\gamma_j\right)X_{j,i}^L+\left(\beta_j+\gamma_j\right)X_{j,i}^U\right]+\eta+\theta.\label{pred_U}
\end{eqnarray}
This is equivalent to minimizing the sum of squared $\delta$-distance in the metric space $\left(\mathcal{K}_{\mathcal{C}}\left(\mathbb{R}\right), \delta\right)$. Theorem \ref{thm:lse-matrix} gives the explicit analytical expression of $\hat{\bdsm{\beta}}$ in matrix form. 
\begin{theorem}\label{thm:lse-matrix}
Consider the linear model (\ref{gen-mod-1})-(\ref{gen-mod-2}), or equivalently, its matrix form (\ref{mod-matrix}). If the data matrix $X$ has full rank, then the least squares estimate $\hat{\bdsm{\beta}}$ defined in (\ref{beta-lse}) is given by
\begin{equation}\label{beta-matrix}
  \hat{\bdsm{\beta}}=\left(\textbf{X}^T\textbf{X}\right)^{-1}\textbf{X}^T\textbf{Y}.
\end{equation}
\end{theorem}
Departing from its matrix form, a series of nice properties of $\hat{\bdsm{\beta}}$ follows immediately from the classical theory of linear models. (See, e.g., Seber (1997).) We summarize them in the following corollaries. 
\begin{corollary}\label{coro:unbias}
$\hat{\bdsm{\beta}}$ in Theorem \ref{thm:lse-matrix} is unbiased.
\end{corollary}
\begin{corollary}\label{coro:consis}
$\hat{\bdsm{\beta}}$ in Theorem \ref{thm:lse-matrix} is consistent.
\end{corollary}
\begin{corollary}\label{coro:cov}
The variance-covariance matrix of $\hat{\bdsm{\beta}}$ is
\begin{equation}\label{beta-cov}
  \text{Cov}\left(\hat{\bdsm{\beta}}\right)=\left(\textbf{X}^T\textbf{X}\right)^{-1}\sigma^2.
\end{equation}
\end{corollary}
\begin{corollary}\label{coro:var-est}
An unbiased estimator of $\sigma^2$ is given by 
\begin{equation}\label{var-est}
  \hat{\sigma^2}=\frac{\left(\textbf{Y}-\textbf{X}\hat{\bdsm{\beta}}\right)^T
	\left(\textbf{Y}-\textbf{X}\hat{\bdsm{\beta}}\right)}{2n-3p-2}.
\end{equation}
\end{corollary}

\section{Positive restrictions}\label{sec:positive}
The model setting requires that $\gamma_j\geq 0$, $\theta\geq 0$, $j=1,\cdots,p$. However, $\hat{\bdsm{\beta}}$ given in (\ref{beta-matrix}) does not automatically guarantee these conditions. In this section, we thoroughly discuss these positive restrictions for the least squares estimation and their implications on the model fitting. We begin by making a few notations and assumptions. 
\begin{notation}
Denote by $X_k^V$ and $Y^V$ the random variables from which $\left\{X_{k,i}^V\right\}_{i=1}^{n}$ and $\left\{Y_i^V\right\}_{i=1}^{n}$ are samples, respectively, where $k=1,\cdots,p$ and $V\in\left\{L, U\right\}$. 
\end{notation}
\begin{notation}
Denote by
\begin{equation*} 
  S_{k,j}=\frac{1}{n}\sum_{i=1}^{n}X_{k,i}^RX_{j,i}^R-\left(\frac{1}{n}\sum_{i=1}^{n}X_{k,i}^R\right)\left(\frac{1}{n}\sum_{i=1}^{n}X_{j,i}^R\right)
\end{equation*}
the sample covariance of $X_k^R$ and $X_j^R$, $k,j=1,\cdots,p$. Similarly, denote by 
\begin{equation*}
  S_k=\frac{1}{n}\sum_{i=1}^{n}X_{k,i}^RY_{k,i}^R-\left(\frac{1}{n}\sum_{i=1}^{n}X_{k,i}^R\right)\left(\frac{1}{n}\sum_{i=1}^{n}Y_i^R\right)
\end{equation*}
the sample covariance of $X_k^R$ and $Y^R$, $k=1,\cdots,p$. 
\end{notation}
\begin{assumption}\label{assumption-1}
The ranges of the predictors $\left\{X_j^R, j=1,\cdots,p\right\}$ are mutually uncorrelated.
\end{assumption}
\begin{assumption}\label{assumption-2}
The range of each predictor $X_j^R$ is empirically positively correlated with the range of the outcome $Y^R$, i.e., $S_j>0$ for $j=1,\cdots,p$. 
\end{assumption}

From the model specification (\ref{gen-mod-1})-(\ref{gen-mod-2}), it is seen that 
\begin{equation}\label{gen-mod-3}
  Y_i^R=\sum_{j=1}^{p}\gamma_jX_{j,i}^R+\theta+\epsilon_i^R,
\end{equation}
where $\epsilon_i^R=\epsilon_i^U-\epsilon_i^L$, $i=1,\cdots, n$. This immediately implies the following results, which give interpretations of the positive parameters $\theta$ and $\gamma_j$, $j=1,\cdots,p$.
\begin{proposition}\label{prop:gamma-theta}
Assume model (\ref{gen-mod-1})-(\ref{gen-mod-2}). Then,
\begin{enumerate}
\item $\begin{bmatrix}\text{Cov}\left(X_1^R,Y^R\right)\\ \cdot\\ \cdot\\ \cdot\\ \text{Cov}\left(X_p^R, Y^R\right)\end{bmatrix}
=\begin{bmatrix}
\text{Cov}\left(X_1^R,X_1^R\right)&\cdot&\cdot&\cdot&\text{Cov}\left(X_1^R,X_p^R\right)\\
\cdot&\cdot&\cdot&\cdot&\cdot\\
\cdot&\cdot&\cdot&\cdot&\cdot\\
\cdot&\cdot&\cdot&\cdot&\cdot\\
\text{Cov}\left(X_p^R,X_1^R\right)&\cdot&\cdot&\cdot&\text{Cov}\left(X_p^R,X_p^R\right)
\end{bmatrix}
\begin{bmatrix}
\gamma_1\\ \cdot\\ \cdot\\ \cdot\\ \gamma_p
\end{bmatrix}$;
\item $\theta=E\left(Y^R\right)-\sum_{j=1}^{p}\gamma_jE\left(X_j^R\right)$. 
\end{enumerate}
\end{proposition}

It can be shown that the positive parameters $\left\{\theta, \gamma_j, j=1,\cdots,p\right\}$ are indeed estimated independently from the rest of the parameters $\left\{\alpha_j, \beta_j, j=1,\cdots, p\right\}$. We list their analytical LS solutions separately in the following theorem.  
\begin{theorem}\label{thm:lse-positivity}
Consider model (\ref{gen-mod-1})-(\ref{gen-mod-2}). Define the sample variance-covariance matrix of $\left\{X_j^R, j=1,\cdots,p\right\}$ as
\begin{equation*}
  \Sigma_{\textbf{X}^R}=\left[\frac{1}{n}\sum_{i=1}^{n}X_{k,i}^RX_{j,i}^R-\left(\frac{1}{n}\sum_{i=1}^{n}X_{k,i}^R\right)\left(\frac{1}{n}\sum_{i=1}^{n}X_{j,i}^R\right)\right]
  _{k,j=1}^{p}:=\left[S_{k,j}\right]_{k,j=1}^{p}.
\end{equation*} 
Additionally, denote by $\Sigma_{\textbf{X}^R,\textbf{Y}^R}$ the vector that contains the sample covariances of $X_k^R$ and $Y^R$, $k=1,\cdots,p$, i.e.
\begin{equation*}
  \Sigma_{\textbf{X}^R,\textbf{Y}^R}=\left[\frac{1}{n}\sum_{i=1}^{n}X_{k,i}^RY_{k,i}^R-\left(\frac{1}{n}\sum_{i=1}^{n}X_{k,i}^R\right)\left(\frac{1}{n}\sum_{}^{}Y_i^R\right)\right]_{k=1}^{p}:=\left[S_k\right]_{k=1}^{p}.
\end{equation*} 
Let $\Gamma=\left[\gamma_1,\cdots,\gamma_p\right]^T$. Then, the LS estimator $\hat{\Gamma}$ is the solution of the linear system
\begin{equation}\label{solu-gamma}  
   \Sigma_{\textbf{X}^R}\Gamma=\Sigma_{\textbf{X}^R,\textbf{Y}^R},
\end{equation}
and the LS estimator of $\theta$ is 
\begin{equation}\label{solu-theta}
  \hat{\theta}=\overline{Y^{R}}-\sum_{j=1}^{p}\hat{\gamma}_j\overline{X_j^R}. 
\end{equation}
\end{theorem}

From (\ref{solu-gamma})-(\ref{solu-theta}) and in view of Proposition \ref{prop:gamma-theta}, we see that $\left\{\hat{\theta}, \hat{\gamma}_j, j=1,\cdots,p\right\}$ are essentially moment estimators of the underlying parameters, which are in fact strongly consistent. This also explains from another perspective the consistency shown in Corollary \ref{coro:consis}. In particular, an important interpretation of Theorem \ref{thm:lse-positivity} is that if at least one of the positive parameters is estimated to be negative for a large sample size, it indicates that the underlying true parameter is negative with a high probability, and forcing the parameter to be positive may result in possible biases. We give a simplified example in the following Corollary to illustrate the implication of the positive restriction on $\left\{\gamma_j,j=1,\cdots,p\right\}$.
\begin{corollary}\label{coro:model-bias}
Consider the univariate model (\ref{uni-mod-1})-(\ref{uni-mod-2}). Let $\hat{\gamma}$ be the LS estimate and $\tilde{\gamma}$ be any constrained LS estimate of $\gamma$ such that $\tilde{\gamma}\geq0$. If $\hat{\gamma}<0$, then
\begin{equation*}
  \sum_{i=1}^{n}\left(Y_i^R-\tilde{Y}_i^R\right)^2\geq
  \sum_{i=1}^{n}\left(Y_i^R-\overline{Y^R}\right)^2,
\end{equation*}
where $\tilde{Y}_i^R$ is the predicted value for $Y_i^R$ based on the constrained LS estimates. The ``='' holds if and only if $\tilde{\gamma}=0$.  
\end{corollary}

For the univariate model, if the LS estimate of $\gamma$ is negative, forcing it to be positive will result in the model being worse than the constant model $\overline{Y^R}$ for the range. Similar biases are expected for the multivariate cases too. Therefore, it is not recommended that a constrained optimization algorithm always be used to ensure positive estimates, if some of the LS estimates $\left\{\hat{\gamma}_j, j=1,\cdots,p\right\}$ are negative. At least a different model that accounts for the negative LS estimates should be considered as an alternative to the constrained linear model. In practice, it is often assumed that the predictors $\left\{X_j, j=1,\cdots,p\right\}$ are independent. We provide a sufficient condition under which the LS estimates $\left\{\hat{\gamma}_j, j=1,\cdots,p\right\}$ are positive with probability converging to one. 
\begin{corollary}\label{coro:lse-positivity}
Under Assumption \ref{assumption-1}, $\hat{\gamma}_j>0$ with probability going to one if Assumption \ref{assumption-2} is met.  
\end{corollary}

Intuitively, under the circumstance of independent predictors, model (\ref{gen-mod-1})-(\ref{gen-mod-2}) implies that 
$$\text{Cov}\left(X_j^R,Y^R\right)>0,\ \ j=1,\cdots,p.$$
Consequently, for data that the model is appropriate for, the sample covariances $\left\{S_j, j=1,\cdots,p\right\}$ are positive almost surely, which by Theorem \ref{thm:lse-positivity} is sufficient to ensure the positiveness of  $\left\{\hat{\gamma}_j, j=1,\cdots,p\right\}$. Otherwise, the $\hat{\gamma}_j$'s can be negative, but that is essentially because one or more of the predictors are negatively correlated with the outcome in range and hence the model is not appropriate. 

From the preceding discussion, if $\hat{\gamma}_j>0, j=1,\cdots,p$, it means that the model fits the linear structure of the data very well. At this point, if $\hat{\theta}<0$, it may not be worth forcing it to be positive using a constrained optimization, as that may bring unnecessary biases. The following theorem gives a guidance of judgment for such a situation.
\begin{theorem}\label{thm:r-positive}
Assume model (\ref{gen-mod-1})-(\ref{gen-mod-2}), or its equivalent matrix form (\ref{mod-matrix}). Let 
\begin{equation}
  \hat{Y}_i^R=\sum_{j=1}^{p}\gamma_jX_{j,i}^R+\theta
\end{equation}
be the model predicted value for $Y_i^R$. Then,
\begin{equation}
  P\left(\hat{Y}_i^R<0\right)
  \leq\frac{Var\left(Y_i^R\right)-Var\left(\hat{Y}_i^R\right)}{\left(Y_i^R\right)^2}
  =\frac{2\sigma^2}{\left(Y_i^R\right)^2}.
\end{equation}
\end{theorem}  

Given a negative $\hat{\theta}$, it is possible to get negative predicts for $Y^R$. However, if the unexplained variance of $Y^R$ is very small compared to the scale of $\left(Y^R\right)^2$, the chance to get a negative predict is tiny, and the rare cases of negative predict, if happened, can be rounded up to $0$. In practice, the unexplained variance of $Y^R$ is estimated by $2\hat{\sigma}^2$, which is then compared to the scale of $\left(Y^R\right)^2$ from the data to decide whether to stay with the negative unbiased LS estimate $\hat{\theta}$ or resort to a constrained LS estimate.

\section{Simulation}
\label{sec:simulation}
We present a simulation study to demonstrate the empirical performance of the LS estimates and compare our model to some peer models in the literature. In particular, we consider the following four model configurations:\\

\begin{itemize}
\item I: p=1, $\eta,\alpha_1,\beta_1\sim\text{Unif}\left(0,4\right)$, $\theta,\gamma_1\sim\text{Unif}\left(1,3\right)$, and $\epsilon_i^L,\epsilon_i^U\sim\text{Unif}\left(0,\sigma^2\right)$ with $\sigma\sim\text{Unif}\left(2,4\right)$, $i=1,\cdots,n$;
\item II: p=1, $\eta,\alpha_1,\beta_1\sim\text{Unif}\left(-4,0\right)$, $\theta,\gamma_1\sim\text{Unif}\left(1,3\right)$, and $\epsilon_i^L,\epsilon_i^U\sim\text{Unif}\left(0,\sigma^2\right)$ with $\sigma\sim\text{Unif}\left(2,4\right)$, $i=1,\cdots,n$;
\item III: p=3, $\eta,\alpha_j,\beta_j\sim\text{Unif}\left(-4,4\right)$, $\theta,\gamma_j\sim\text{Unif}\left(1,3\right)$, $j=1,2,3$, and $\epsilon_i^L,\epsilon_i^U\sim\text{Unif}\left(0,\sigma^2\right)$ with $\sigma\sim\text{Unif}\left(2,4\right)$, $i=1,\cdots,n$.
\end{itemize} 

\begin{figure}[ht]
\centering
\includegraphics[ height=2.000in, width=2.200in]{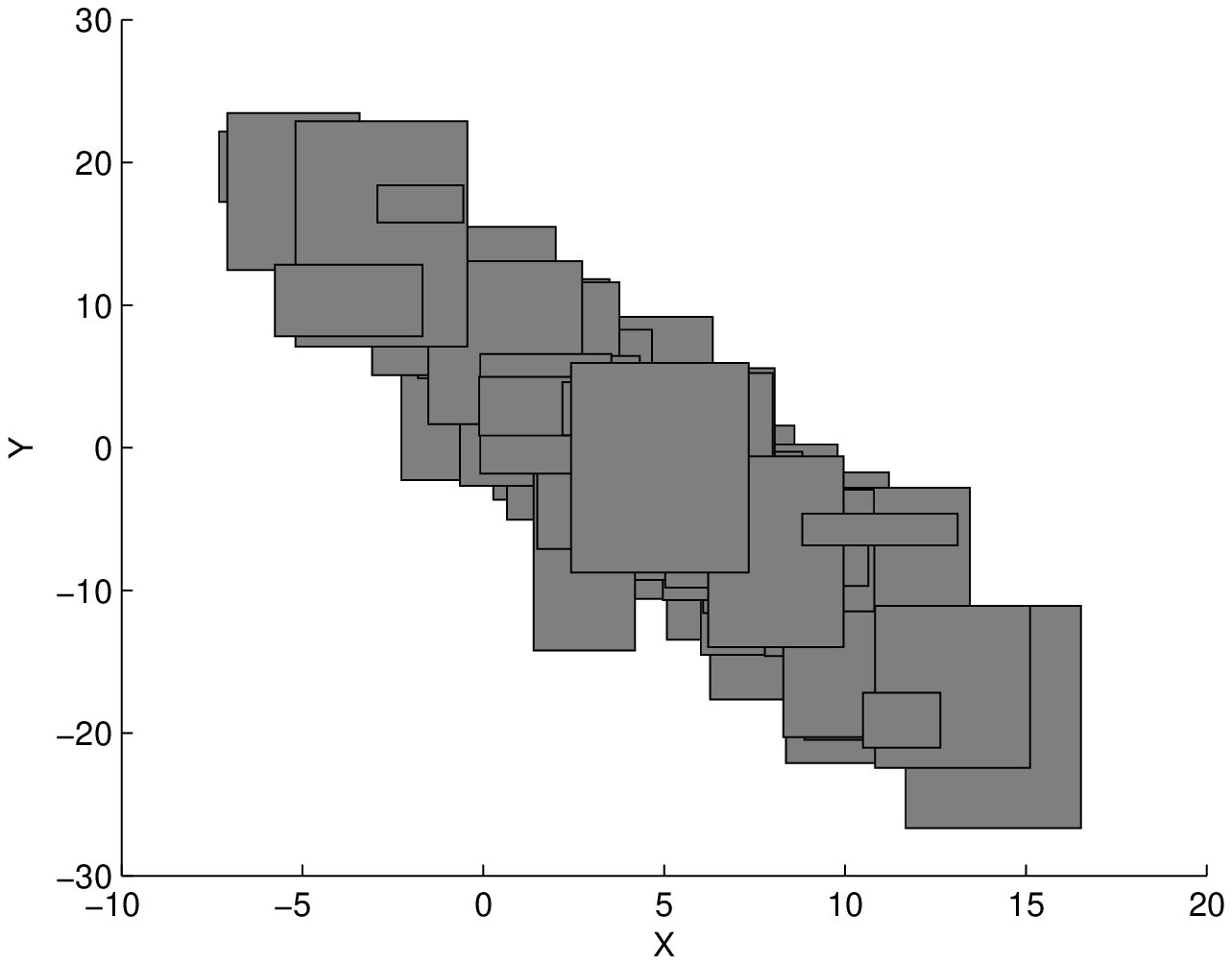}
\includegraphics[ height=2.000in, width=2.200in]{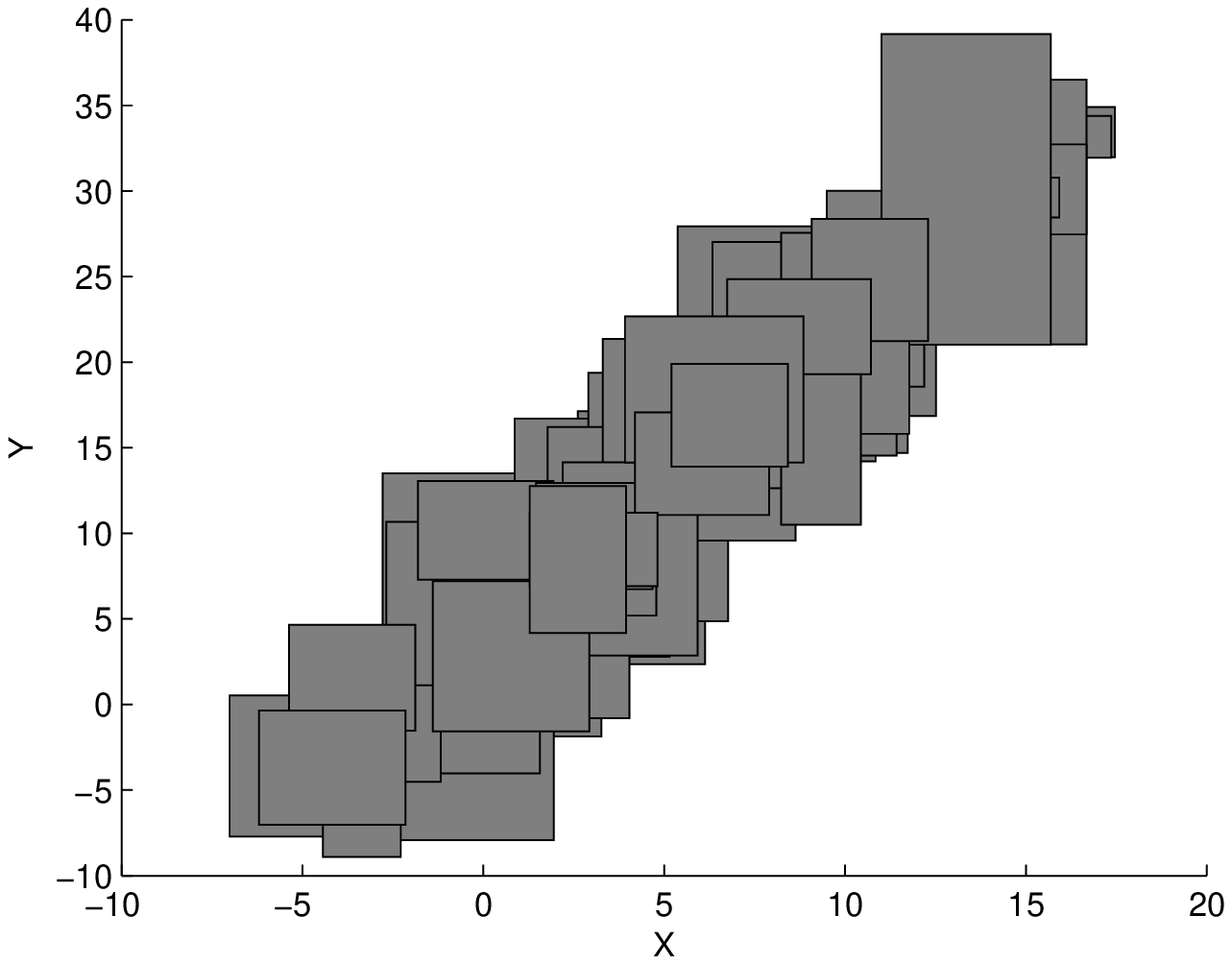}
\caption{Plots of simulated data from model I and II, respectively, each with sample size $n=100$.}
\label{fig:sim-data-1-2}
\end{figure}

\begin{figure}[ht]
\centering
\includegraphics[ height=2.000in, width=2.200in]{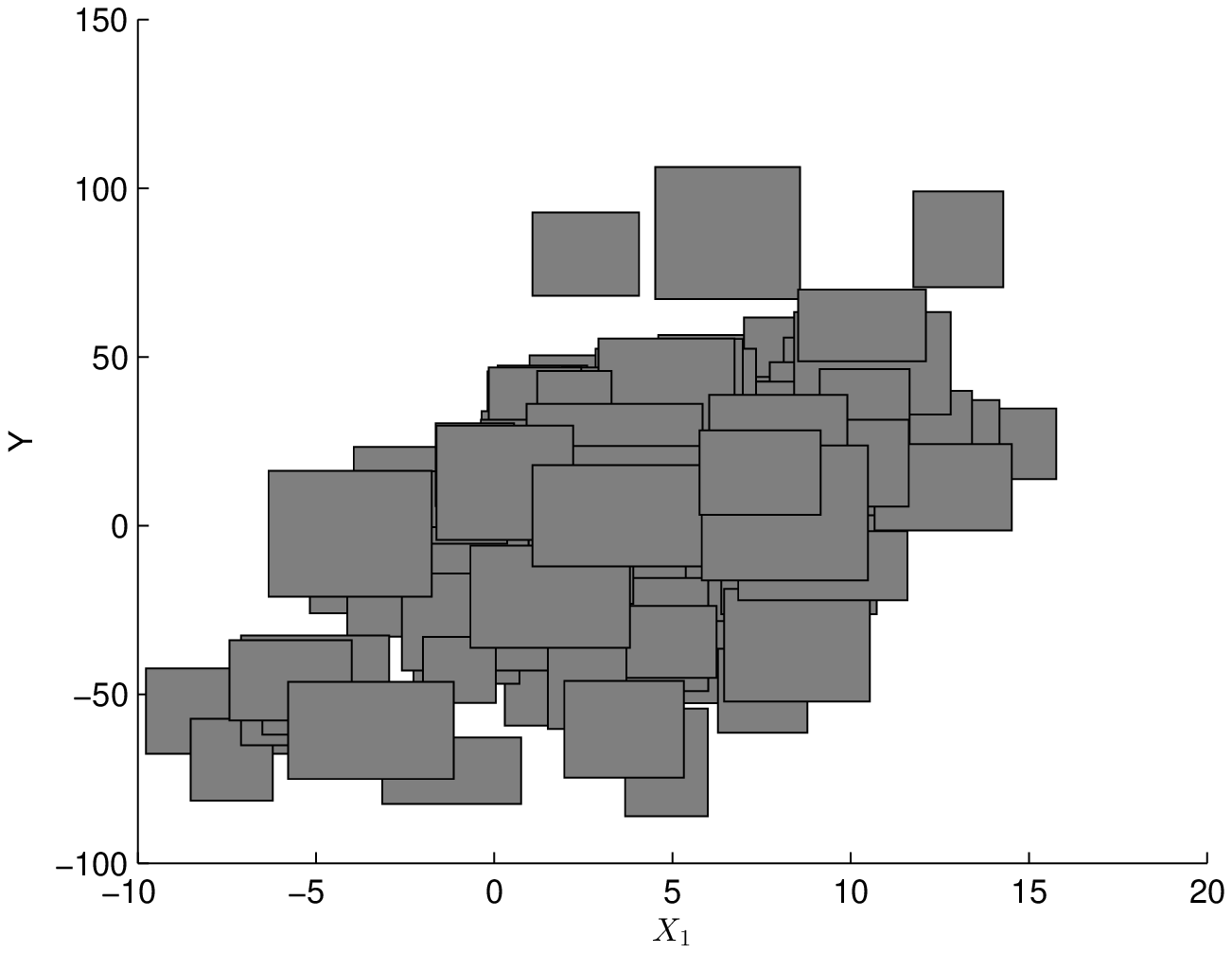}
\includegraphics[ height=2.000in, width=2.200in]{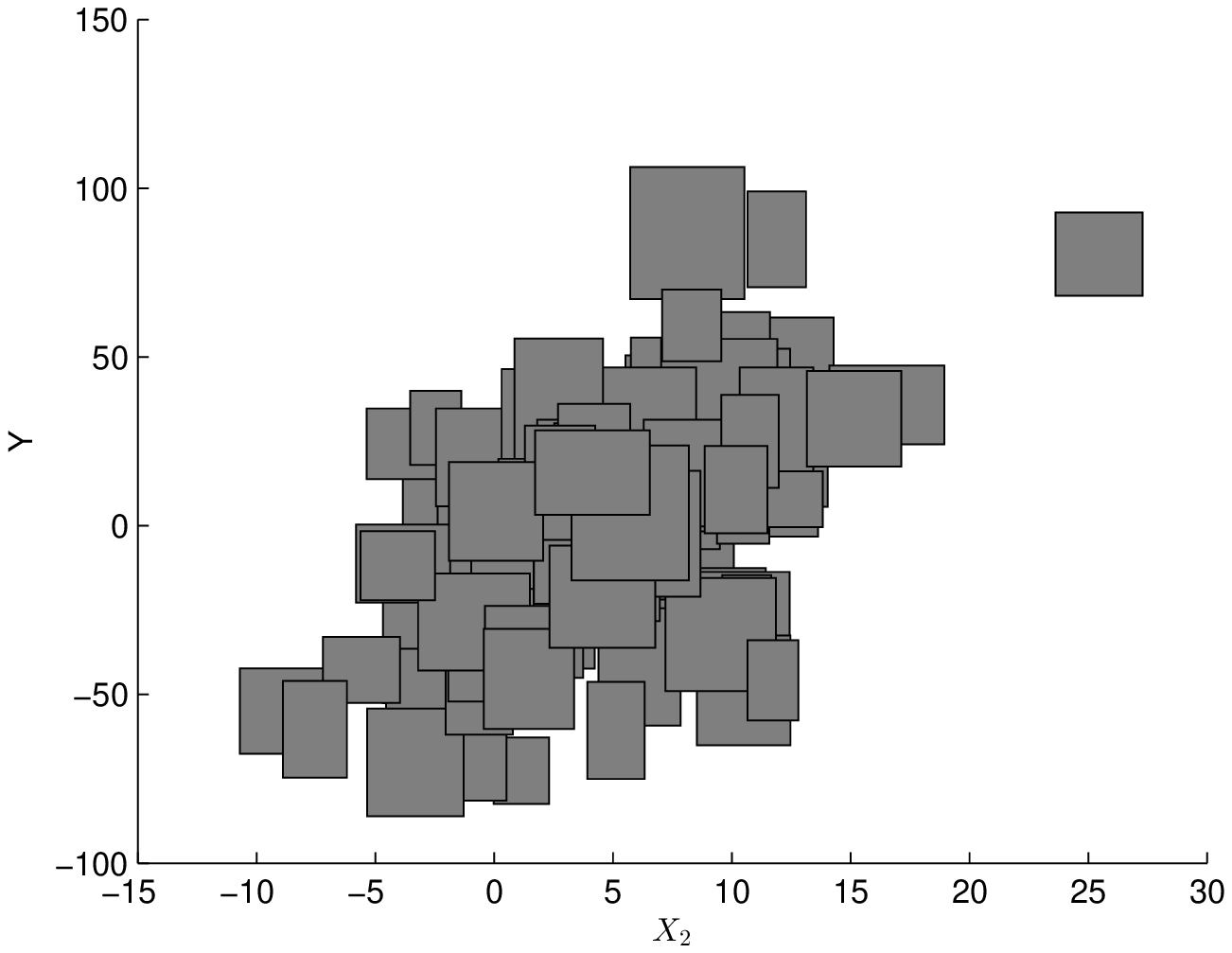}\\
\includegraphics[ height=2.000in, width=2.200in]{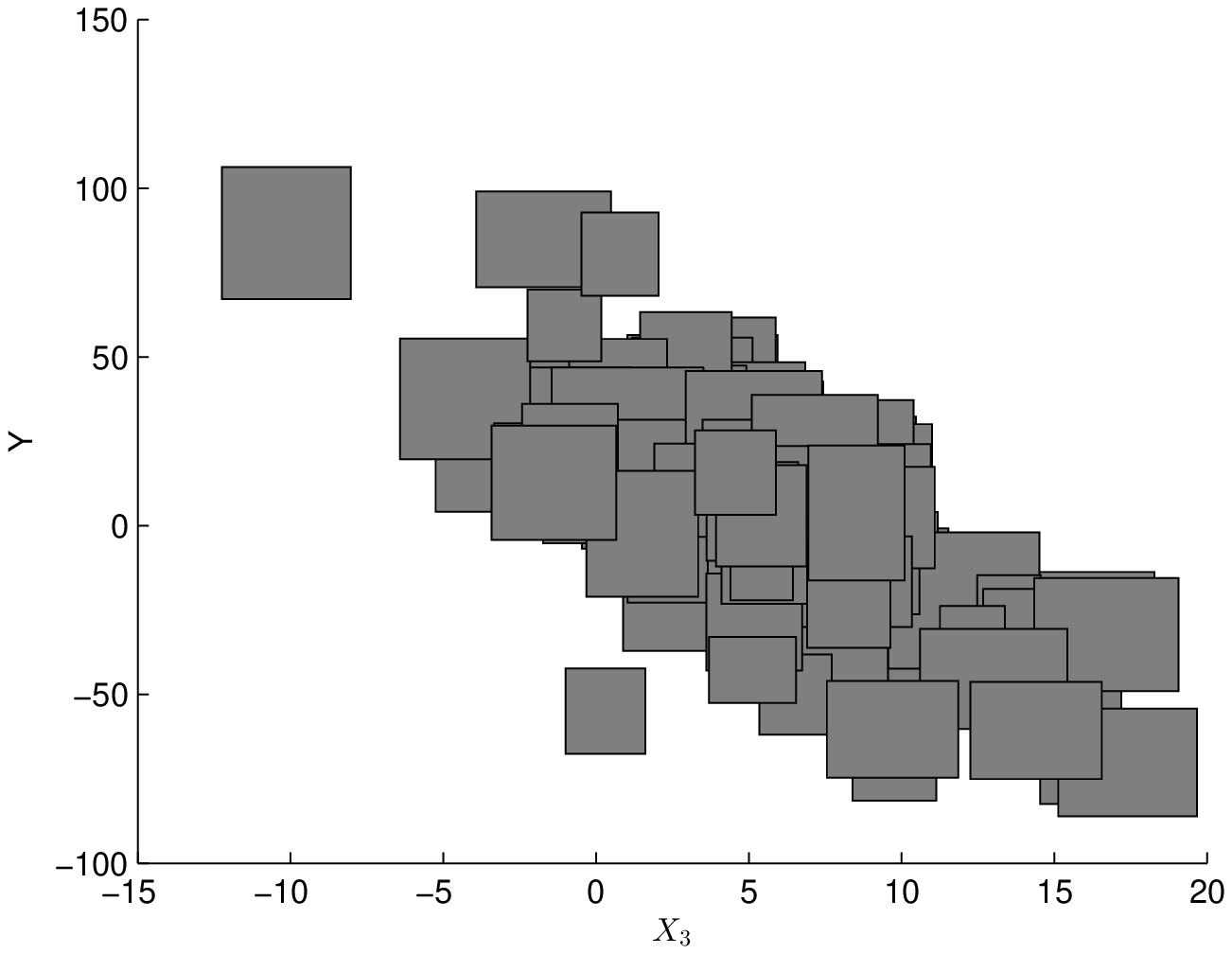}\\
\caption{Plots of $Y$ against $X_1$, $X_2$, and $X_3$, respectively, of a simulated data from model III with sample size $n=100$.}
\label{fig:sim-data-3}
\end{figure}

The first two are univariate models, with positive and negative interval correlations between $Y$ and $X_1$, respectively. Figure \ref{fig:sim-data-1-2} shows a plot of simulated data with $n=100$ observations from each of the two models.  The third one is a 3-dimensional model, with $Y$ and $X_j$, $j=1,2,3$, either positively or negatively correlated. A particular data with $n=100$ observations simulated from this model is visualized in Figure \ref{fig:sim-data-3}, where it is seen that $Y$ is positively correlated with both $X_1$ and $X_2$, and negatively correlated with $X_3$. 

To investigate the empirical performance of the LS estimation, we simulate 500 independent data from each of the three model configurations and calculate the LS estimates of the parameters for each simulated data. The results are summarized into Table \ref{tab:sim}. The mean relative error (MRE) for the estimated coefficient matrix $\hat{\bdsm{\beta}}$ and variance of error $\hat{\sigma^2}$, given a fixed sample size $n$, are defined as
\begin{equation*}
  \text{MRE}\left(\hat{\bdsm{\beta}}\right)=\frac{1}{500}\sum_{k=1}^{500}
	\frac{\left\|\hat{\bdsm{\beta}}_{k}-\bdsm{\beta}_k\right\|}{\left\|\bdsm{\beta}_k\right\|},
\end{equation*}
where $\left\|\cdot\right\|$ denotes the Euclidean norm, and 
\begin{equation*}
  \text{MRE}\left(\hat{\sigma^2}\right)=\frac{1}{500}\sum_{k=1}^{500}
	\frac{|\hat{\sigma^2}_k-\sigma^2_k|}{\sigma^2_k},
\end{equation*}
respectively. We simulate observations for each $X_i$ independently, so Assumption \ref{assumption-1} is automatically satisfied. Assumption \ref{assumption-2} is checked before we compute the LS estimates for the parameters for each simulated data. If it is satisfied, then $\hat{\bdsm{\beta}}$ is calculated by (\ref{beta-matrix}), which according to Corollary \ref{coro:lse-positivity} produces positive $\hat{\gamma_j}$, $j=1,\cdots,p$ with probability going to one. If otherwise Assumption \ref{assumption-2} is violated, a constrained optimization algorithm is employed to calculate $\hat{\bdsm{\beta}}$, with the constraints that $\hat{\gamma_j}\geq 0$, $j=1,\cdots,p$ and $\hat{\theta}\geq 0$. For this paper, we have used the Matlab function $fmincon.m$ to compute the constrained LS estimates. Consistent to our theorems, we see that the MRE's for both $\hat{\bdsm{\beta}}$ and $\hat{\sigma^2}$ converge to $0$ as sample size increases. Especially, if the model really fits the data, which is the case for our simulation, the unconstrained LS estimate given in (\ref{beta-matrix}) is sufficient, without the need of a constrained optimization algorithm, with probability going to one.  

\begin{table}[htbp]
  \centering
  \caption{Evaluation of the LS estimation for simulated data based on 500 independent repetitions.}
  \bigskip
    \begin{tabular}{lrrrrr}
    \toprule
          &       & \multicolumn{1}{c}{\textbf{}} & \multicolumn{1}{c}{\textbf{}} & \multicolumn{1}{c}{\textbf{}} & \multicolumn{1}{c}{\textbf{}} \\
    \midrule
    \textbf{} & \multicolumn{1}{c}{\textbf{n}} & \multicolumn{1}{c}{\textbf{MRE ($\hat{\bdsm{\beta}}$)}} & \multicolumn{1}{c}{\textbf{MRE ($\hat{\sigma^2}$)}} & \multicolumn{1}{c}{\textbf{Unconstrained}} & \multicolumn{1}{c}{\textbf{Constrained}} \\
    \textbf{} & \multicolumn{1}{c}{\textbf{}} & \multicolumn{1}{c}{\textbf{}} & \multicolumn{1}{c}{\textbf{}} & \multicolumn{1}{c}{\textbf{}} & \multicolumn{1}{c}{\textbf{}} \\
    \textbf{Model I} & \multicolumn{1}{c}{100} & 0.4764 & 0.0982 & 496   & 4 \\
    \textbf{} & \multicolumn{1}{c}{200} & 0.3806 & 0.0796 & 499   & 1 \\
    \textbf{} & \multicolumn{1}{c}{300} & 0.3539 & 0.0726 & 500   & 0 \\
    \textbf{} & \multicolumn{1}{c}{400} & 0.3386 & 0.0695 & 500   & 0 \\
    \textbf{} & \multicolumn{1}{c}{} &       &       &       &  \\
    \textbf{Model II} & \multicolumn{1}{c}{100} & 0.4468 & 0.0985 & 499   & 1 \\
    \textbf{} & \multicolumn{1}{c}{200} & 0.3844 & 0.086 & 499   & 1 \\
    \textbf{} & \multicolumn{1}{c}{300} & 0.3613 & 0.0764 & 500   & 0 \\
    \textbf{} & \multicolumn{1}{c}{400} & 0.3331 & 0.0728 & 500   & 0 \\
    \textbf{} & \multicolumn{1}{c}{} &       &       &       &  \\
    \textbf{Model III} & \multicolumn{1}{c}{100} & 0.4581 & 0.0832 & 460   & 40 \\
    \textbf{} & \multicolumn{1}{c}{200} & 0.3305 & 0.0546 & 473   & 27 \\
    \textbf{} & \multicolumn{1}{c}{300} & 0.2666 & 0.0463 & 478   & 22 \\
    \textbf{} & \multicolumn{1}{c}{400} & 0.2352 & 0.0391 & 481   & 19 \\
    \bottomrule
    \end{tabular}%
  \label{tab:sim}%
\end{table}%

Next, we carry out more delicate investigations into the parameter estimation with a particular model randomly generated from configuration III. The exact parameter values are listed in the second column of Table \ref{tab:sim_1model}. We simulate a random sample of size $n=300$ from this model and estimate the coefficient matrix $\bdsm{\beta}$ using the algorithm described in the preceding paragraph. The procedure is repeated for 500 times independently, and the mean estimates and mean variances are reported in columns 3 and 4, respectively. It is seen that the mean estimates are very close to the corresponding true values. The empirical variances of these estimates for the 500 repetitions are displayed in column 5, which are satisfactorily close to the calculated variances in column 4. 

\begin{table}[htbp]
  \centering
  \caption{Parameter estimation for one particular model based on 500 independent repetitions.}
  \bigskip
    \begin{tabular}{rrrrr}
    \toprule
    \multicolumn{1}{c}{\textbf{Parameter}} & \multicolumn{1}{c}{\textbf{True Value}} & \multicolumn{1}{c}{\textbf{Mean Estimate}} & \multicolumn{1}{c}{\textbf{Estimated Variance}} & \multicolumn{1}{c}{\textbf{Empirical Variance}} \\
    \midrule
          &       &       &       &  \\
    $\eta$ & 1.4932 & 1.4499 & 2.0048 & 1.8581 \\
    $\alpha_1$ & 1.6419 & 1.6457 & 0.0520 & 0.0527 \\
    $\beta_1$ & 1.5542 & 1.5527 & 0.0521 & 0.0524 \\
    $\alpha_2$ & -1.8902 & -1.9098 & 0.0521 & 0.0557 \\
    $\beta_2$ & -3.2780 & -3.2585 & 0.0521 & 0.0558 \\
    $\alpha_3$ & -2.4036 & -2.3967 & 0.0518 & 0.0519 \\
    $\beta_3$ & -1.8451 & -1.8528 & 0.0518 & 0.0508 \\
    $\theta$ & 1.7999 & 1.8149 & 3.8753 & 3.5164 \\
    $\gamma_1$ & 1.2086 & 1.2276 & 0.1033 & 0.1031 \\
    $\gamma_2$ & 2.5633 & 2.5347 & 0.1035 & 0.1112 \\
    $\gamma_3$ & 2.5436 & 2.5477 & 0.1028 & 0.0971 \\
    \bottomrule
    \end{tabular}%
  \label{tab:sim_1model}%
\end{table}%

Finally, we compare our linear model to the M model by Blanco-Fern\'andez et al. (2011) and the constrained center and range method (CCRM) by Lima Neto and De Carvalho (2010). We presented in Section \ref{sec:model-compare} that these two models are essentially reduced forms of our model. Here we give empirical evidence based on their predicting performances. We simulate 500 independent samples from Model I, II, and III, with training sample size $n=60, 100, 200, 300$, respectively. For each sample, we simulate another $n/4$ observations as the validation set. We use the mean squared error (MSE) of the center, radius (half-range), and the interval as a whole, for the validation set as our measures of predicting performance. Specifically, they are defined as
\begin{eqnarray*}
  \text{MSEC}&=&\frac{4}{n}\sum_{i=1}^{n/4}\left(\hat{Y}^c_i-Y^c_i\right)^2,\\
  \text{MSER}&=&\frac{4}{n}\sum_{i=1}^{n/4}\left(\hat{Y}^r_i-Y^r_i\right)^2,\\
  \text{MSEI}&=&\text{MSEC}+\text{MSER}.
\end{eqnarray*}
The LS solution for the parameters of the M model is calculated according to the formulas given in Blanco-Fern\'andez et al. (2011). The CCRM is implemented using the R function $ccrm$ in the $iRegression$ package. The M model was only developed for the univariate case, so it is excluded in the multiple case (Model III).  Numerical results for comparing the three methods are shown in Table \ref{tab:sim_compare}. Just as we expected, the performance of our model is consistently significantly better than the other two models across different model configurations and sample sizes. Especially for Model III, the average MSEC of the CCRM is about 3 times bigger than that of our model, which results from the increased number of predictors. That is, the expanded flexibility of our model increases proportionally with the size of the model. The more predictors we include in the model, the more increased flexibility we have over the CCRM and the M model. 

\begin{table}[htbp]
  \centering
  \caption{Comparison of CCRM and our model for simulated data based on the average of 500 independent repetitions.}
  \bigskip
    \begin{tabular}{rrrrrrrrrrr}
    \toprule
          &       & \multicolumn{3}{c}{\textbf{M}} & \multicolumn{3}{c}{\textbf{CCRM}} & \multicolumn{3}{c}{\textbf{Our Linear Model}} \\
    \midrule
          &       & \multicolumn{9}{c}{\textbf{}} \\
    \multicolumn{1}{c}{} & \multicolumn{1}{c}{n} & \multicolumn{1}{c}{MSEC} & \multicolumn{1}{c}{MSER} & \multicolumn{1}{c}{\textbf{MSEI}} & \multicolumn{1}{c}{MSEC} & \multicolumn{1}{c}{MSER} & \multicolumn{1}{c}{\textbf{MSEI}} & \multicolumn{1}{c}{MSEC} & \multicolumn{1}{c}{MSER} & \multicolumn{1}{c}{\textbf{MSEI}} \\
    \multicolumn{1}{c}{} & \multicolumn{1}{c}{} & \multicolumn{1}{c}{} & \multicolumn{1}{c}{} & \multicolumn{1}{c}{} & \multicolumn{1}{c}{} & \multicolumn{1}{c}{} & \multicolumn{1}{c}{} & \multicolumn{1}{c}{} & \multicolumn{1}{c}{} & \multicolumn{1}{c}{\textbf{}} \\
    \multicolumn{1}{c}{\textbf{Model I}} & \multicolumn{1}{l}{60} & \multicolumn{1}{l}{6.3398} & \multicolumn{1}{l}{4.6962} & \multicolumn{1}{l}{\textbf{11.0361}} & \multicolumn{1}{l}{6.2085} & \multicolumn{1}{l}{4.8585} & \multicolumn{1}{l}{\textbf{11.067}} & \multicolumn{1}{l}{4.8214} & \multicolumn{1}{l}{4.1395} & \multicolumn{1}{l}{\textbf{8.961}} \\
    \multicolumn{1}{c}{} & \multicolumn{1}{l}{100} & \multicolumn{1}{l}{6.1048} & \multicolumn{1}{l}{4.4821} & \multicolumn{1}{l}{\textbf{10.5869}} & \multicolumn{1}{l}{6.1318} & \multicolumn{1}{l}{4.7239} & \multicolumn{1}{l}{\textbf{10.8557}} & \multicolumn{1}{l}{4.775} & \multicolumn{1}{l}{3.9936} & \multicolumn{1}{l}{\textbf{8.7686}} \\
    \multicolumn{1}{c}{} & \multicolumn{1}{l}{200} & \multicolumn{1}{l}{6.1018} & \multicolumn{1}{l}{4.6074} & \multicolumn{1}{l}{\textbf{10.7092}} & \multicolumn{1}{l}{6.0296} & \multicolumn{1}{l}{4.5987} & \multicolumn{1}{l}{\textbf{10.6283}} & \multicolumn{1}{l}{4.8072} & \multicolumn{1}{l}{4.0166} & \multicolumn{1}{l}{\textbf{8.8238}} \\
    \multicolumn{1}{c}{} & \multicolumn{1}{l}{300} & \multicolumn{1}{l}{5.991} & \multicolumn{1}{l}{4.494} & \multicolumn{1}{l}{\textbf{10.4849}} & \multicolumn{1}{l}{5.9705} & \multicolumn{1}{l}{4.7316} & \multicolumn{1}{l}{\textbf{10.7021}} & \multicolumn{1}{l}{4.6907} & \multicolumn{1}{l}{3.9587} & \multicolumn{1}{l}{\textbf{8.6493}} \\
    \multicolumn{1}{c}{} & \multicolumn{1}{l}{} &       &       & \textbf{} &       &       & \textbf{} &       &       &  \\
    \multicolumn{1}{c}{\textbf{Model II}} & \multicolumn{1}{l}{60} & \multicolumn{1}{l}{6.5056} & \multicolumn{1}{l}{4.4005} & \multicolumn{1}{l}{\textbf{10.9062}} & \multicolumn{1}{l}{6.4052} & \multicolumn{1}{l}{4.8404} & \multicolumn{1}{l}{\textbf{11.2456}} & \multicolumn{1}{l}{4.9052} & \multicolumn{1}{l}{4.0148} & \multicolumn{1}{l}{\textbf{8.92}} \\
    \multicolumn{1}{c}{} & \multicolumn{1}{l}{100} & \multicolumn{1}{l}{6.178} & \multicolumn{1}{l}{4.567} & \multicolumn{1}{l}{\textbf{10.745}} & \multicolumn{1}{l}{6.0754} & \multicolumn{1}{l}{4.6265} & \multicolumn{1}{l}{\textbf{10.7019}} & \multicolumn{1}{l}{4.7378} & \multicolumn{1}{l}{3.9721} & \multicolumn{1}{l}{\textbf{8.7098}} \\
    \multicolumn{1}{c}{} & \multicolumn{1}{l}{200} & \multicolumn{1}{l}{6.1372} & \multicolumn{1}{l}{4.6152} & \multicolumn{1}{l}{\textbf{10.7525}} & \multicolumn{1}{l}{6.0185} & \multicolumn{1}{l}{4.6246} & \multicolumn{1}{l}{\textbf{10.6431}} & \multicolumn{1}{l}{4.6802} & \multicolumn{1}{l}{3.9186} & \multicolumn{1}{l}{\textbf{8.5988}} \\
    \multicolumn{1}{c}{} & \multicolumn{1}{l}{300} & \multicolumn{1}{l}{6.0026} & \multicolumn{1}{l}{4.4992} & \multicolumn{1}{l}{\textbf{10.5018}} & \multicolumn{1}{l}{5.9019} & \multicolumn{1}{l}{4.7009} & \multicolumn{1}{l}{\textbf{10.6029}} & \multicolumn{1}{l}{4.5837} & \multicolumn{1}{l}{3.9066} & \multicolumn{1}{l}{\textbf{8.4903}} \\
    \multicolumn{1}{c}{} & \multicolumn{1}{l}{} &       &       &       &       &       & \textbf{} &       &       &  \\
    \multicolumn{1}{c}{\textbf{Model III}} & \multicolumn{1}{l}{60} & \multicolumn{1}{c}{-} & \multicolumn{1}{c}{-} & \multicolumn{1}{c}{-} & \multicolumn{1}{l}{14.1623} & \multicolumn{1}{l}{5.1865} & \multicolumn{1}{l}{\textbf{19.3488}} & \multicolumn{1}{l}{5.1949} & \multicolumn{1}{l}{4.9172} & \multicolumn{1}{l}{\textbf{10.1122}} \\
    \multicolumn{1}{c}{} & \multicolumn{1}{l}{100} & \multicolumn{1}{c}{-} & \multicolumn{1}{c}{-} & \multicolumn{1}{c}{-} & \multicolumn{1}{l}{13.2387} & \multicolumn{1}{l}{4.8486} & \multicolumn{1}{l}{\textbf{18.0873}} & \multicolumn{1}{l}{5.0919} & \multicolumn{1}{l}{4.8285} & \multicolumn{1}{l}{\textbf{9.9205}} \\
    \multicolumn{1}{c}{} & \multicolumn{1}{l}{200} & \multicolumn{1}{c}{-} & \multicolumn{1}{c}{-} & \multicolumn{1}{c}{-} & \multicolumn{1}{l}{13.3159} & \multicolumn{1}{l}{4.7531} & \multicolumn{1}{l}{\textbf{18.069}} & \multicolumn{1}{l}{4.7472} & \multicolumn{1}{l}{4.6946} & \multicolumn{1}{l}{\textbf{9.4418}} \\
    \multicolumn{1}{c}{} & \multicolumn{1}{l}{300} & \multicolumn{1}{c}{-} & \multicolumn{1}{c}{-} & \multicolumn{1}{c}{-} & \multicolumn{1}{l}{13.1125} & \multicolumn{1}{l}{4.825} & \multicolumn{1}{l}{\textbf{17.9375}} & \multicolumn{1}{l}{4.6887} & \multicolumn{1}{l}{4.634} & \multicolumn{1}{l}{\textbf{9.3227}} \\
    \bottomrule
    \end{tabular}%
  \label{tab:sim_compare}%
\end{table}%

\section{A real data application}
\begin{figure}[ht]
\centering
\includegraphics[ height=2.000in, width=2.200in]{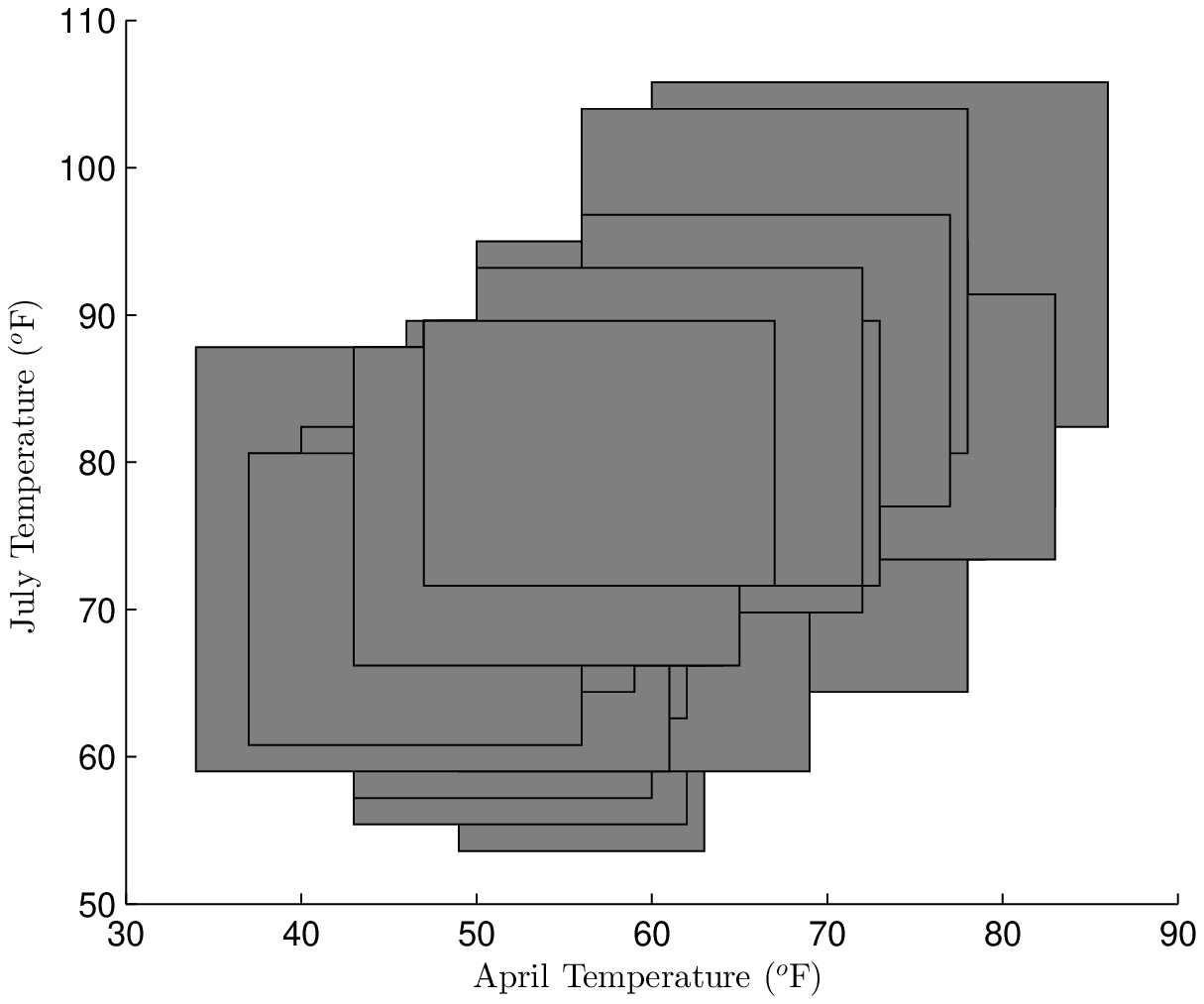}
\includegraphics[ height=2.000in, width=2.200in]{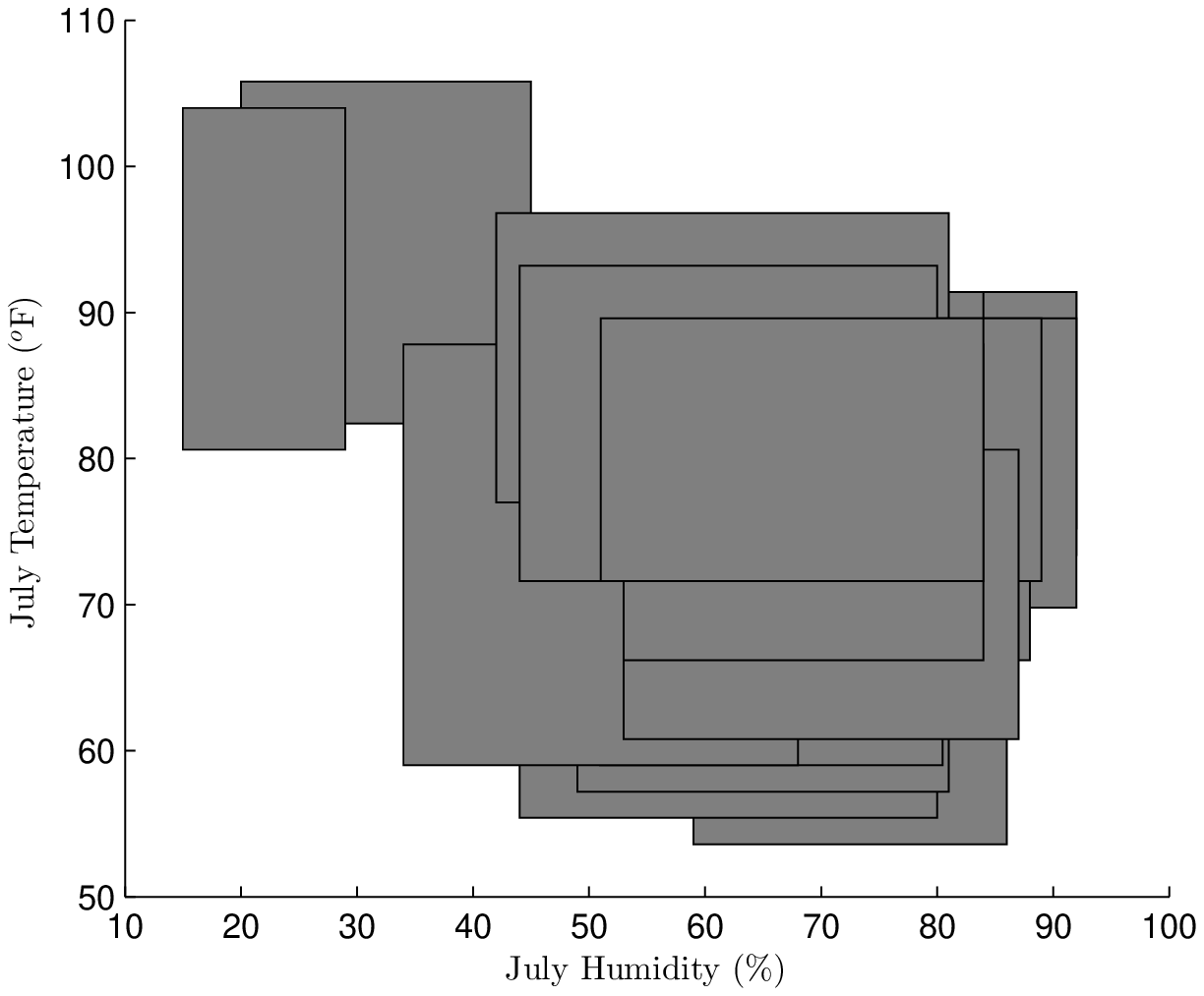}
\caption{Left: plot of July temperature versus April temperature. Right: plot of July temperature versus July relative humidity.}
\label{fig:real-data}
\end{figure}

In this section, we apply our linear model to analyze an interval-valued climate data provided by the National Oceanic and Atmospheric Administration (NOAA) and publicly available. The data contains three variables. The outcome variable, which we denote by $Y$, is the average [minimum, maximum] temperature in July based on weather data collected from 1981 to 2010 by the NOAA National Climatic Data Center of the United States. The first predictor $X_1$ is the corresponding average temperature range in April. The second predictor $X_2$ is the [morning, afternoon] relative humidity in July averaged for the years 1961 to 1990. Relative humidity measures the actual amount of moisture in the air as a percentage of the maximum amount of moisture the air can hold, and it corresponds negatively to the temperature. All the three interval-valued variables are observed for 51 large US cities. By this analysis, we aim to model the summer (July) temperature by an affine function of the spring (April) temperature and the July relative humidity. We randomly split the full data into a training set of 40 observations and a validation set of 11 observations.  Figure \ref{fig:real-data} plots $Y$ against $X_1$ and $X_2$, respectively, for the training set. It is checked that 1) the data matrix $\bdsm{X}$ has full rank; 2) the sample correlation of $X_1^R$ and $X_2^R$ has a p-value greater than $0.05$; 3) the sample correlations of $X_i^R$ and $Y^R$, $i=1,2$, are both positive. So all the theoretical results we developed in Section \ref{sec:para-esti} and \ref{sec:positive} should apply. This means that with large probability we can get the LS estimates $\hat{\bdsm{\beta}}$ simply by formula (\ref{beta-matrix}), without any constrained optimization algorithm. The LS estimates of the parameters are found to be
\begin{eqnarray*}
  &&\eta=45.7740,\ \theta=1.5419,\\
	&&\alpha_1=0.2069,\ \beta_1=0.3392,\ \gamma_1=0.8839,\\
	&&\alpha_2=-0.0593,\ \beta_2=-0.0948,\ \gamma_2=0.0234,
\end{eqnarray*}
and the estimated variance of residual according to Corollary \ref{coro:var-est} is
\begin{equation*}
  \hat{\sigma^2}=16.0454.
\end{equation*}
It follows that the fitted linear model is 
\begin{eqnarray*}
  Y^L&=&45.7740+0.2069X_1^L+0.3392X_1^U-0.0593X_2^L-0.0948X_2^U+\epsilon^L,\\
  Y^U&=&47.3159-0.6269X_1^L+1.1731X_1^U-0.0826X_2^L-0.0714X_2^U+\epsilon^U,
\end{eqnarray*}
where $\epsilon^L, \epsilon^U$ are i.i.d. random variables with mean $0$ and variance $16.0454$. For comparison purposes, we also fit a CCRM model to the data, which turns out to be
\begin{eqnarray*}
  Y^C&=&57.2428+0.5768X_1^C-0.1921X_2^C+\epsilon^C,\\
  Y^R&=&1.5419+0.8339X_1^R+0.0234X_2^R+\epsilon^R,
\end{eqnarray*}
where the random errors $\epsilon^C, \epsilon^R$ have both means $0$, and variances $19.7509$ and $12.9017$, respectively. The predicting performances on the validation set of both models are reported in table \ref{tab:real-data}. Consistent with our theoretical analysis in Section \ref{sec:model-compare} and our simulation study in Section \ref{sec:simulation}, our linear model has much more flexibility than the existing reduced models such as CCRM, which leads to the much improved predicting performance even for a small data set as presented here. 

\begin{table}[htbp]
  \centering
  \caption{Predicting performance comparison of CCRM and our model for the real data.}
  \bigskip
    \begin{tabular}{rrrr}
    \toprule
          & \multicolumn{1}{c}{\textbf{MSEC}} & \multicolumn{1}{c}{\textbf{MSER}} & \multicolumn{1}{c}{\textbf{MSEI}} \\
    \midrule
          & \multicolumn{1}{c}{\textbf{}} & \multicolumn{1}{c}{\textbf{}} & \multicolumn{1}{c}{\textbf{}} \\
    \textbf{Our Linear Model} & \multicolumn{1}{l}{8.7484} & \multicolumn{1}{l}{4.1004} & \multicolumn{1}{l}{12.8488} \\
    \textbf{CCRM} & \multicolumn{1}{l}{10.8631} & \multicolumn{1}{l}{4.1004} & \multicolumn{1}{l}{14.9634} \\
    \bottomrule
    \end{tabular}%
  \label{tab:real-data}%
\end{table}%

\section{Conclusion}
We have introduced a linear model for interval-valued data based on the affine operators in the cone $\mathcal{C} = \{ (x, y) \in \mathbb{R}^2 | x \leq y\}$. The new model is shown both theoretically and empirically to have improved flexibility over the existing models in the literature. We present the general model for multiple predictors in matrix form, from which the LS estimators of the model parameters are immediately derived with a series of nice properties from the classical theory of linear models. Some parameters have positive constraints, which we show are closely related to the intrinsic structure of the model. Therefore, it is not recommended to blindly force these parameters to be positive with a constrained optimization algorithm. Instead, it is better to let the data speak for itself by the unconstrained LS estimates and decide later whether to employ a constrained optimization algorithm or resort to a different model, according to the guideline we have provided in the paper. 

\appendix
\section{Proofs}
\subsection{Proof of Proposition \ref{prop:gamma-theta}}
\begin{proof}
From (\ref{gen-mod-3}), since $\epsilon^R$ and $X_k^R$ are uncorrelated,  
\begin{equation*}
  \text{Cov}\left(X_k^R,Y^R\right)
	=\text{Cov}\left(X_k^R,\sum_{j=1}^{p}\gamma_jX_j^R+\theta+\epsilon^R\right)
	=\sum_{j=1}^{p}\gamma_j\text{Cov}\left(X_k^R,X_j^R\right),\ \ k=1,\cdots,p,
\end{equation*}
from which the first result follows. Taking expectations on both sides of (\ref{gen-mod-3}) yields the second result. 
\end{proof}

\subsection{Proof of Theorem \ref{thm:lse-positivity}}
\begin{proof}
Differentiating $\sum_{i=1}^{n}\left[\left(Y_i^L-\hat{Y_i^L}\right)^2+\left(Y_i^U-\hat{Y_i^U}\right)^2\right]$ with respect to $\left\{\alpha_k, \beta_k, \gamma_k, \eta, \theta, k=1,\cdots,p\right\}$, we obtain the system of equations
\begin{eqnarray}
  &&\sum_{i=1}^{n}\left(Y_i^U-\hat{Y_i^U}\right)=0,\label{eqn-1}\\
  &&\sum_{i=1}^{n}\left(Y_i^L-\hat{Y_i^L}\right)=0,\label{eqn-2}\\
  &&\sum_{i=1}^{n}X_{k,i}^R\left(Y_i^U-\hat{Y_i^U}\right)=0,\label{eqn-3}\\
  &&\sum_{i=1}^{n}X_{k,i}^R\left(Y_i^L-\hat{Y_i^L}\right)=0,\label{eqn-4}\\
  &&\sum_{i=1}^{n}X_{k,i}^R\left[\left(Y_i^U-\hat{Y_i^U}\right)+\left(Y_i^L-\hat{Y_i^L}\right)\right]=0,\label{eqn-5}\\
  &&k=1,\cdots,p.\nonumber
\end{eqnarray}
Equations (\ref{eqn-1})-(\ref{eqn-2}) yield 
\begin{equation}\label{eqn-6}
  \sum_{i=1}^{n}Y_i^R=\sum_{j=1}^{p}\gamma_j\left(\sum_{i=1}^{n}X_{j,i}^R\right)+n\theta. 
\end{equation}
Meanwhile, equations (\ref{eqn-3})-(\ref{eqn-4}) yield
\begin{equation}\label{eqn-7}
  \sum_{j=1}^{p}\gamma_j\left(\sum_{i=1}^{n}X_{k,i}^RX_{j,i}^R\right)+\theta\sum_{i=1}^{n}X_{k,i}^R=\sum_{i=1}^{n}X_{k,i}^RY_i^R,\ \  k=1,\cdots,p.
\end{equation}
Plugging (\ref{eqn-6}) into (\ref{eqn-7}), we obtain
\begin{eqnarray}
  &&\sum_{j=1}^{p}\gamma_j\left[\frac{1}{n}\sum_{i=1}^{n}X_{k,i}^RX_{j,i}^R-\left(\frac{1}{n}\sum_{i=1}^{n}X_{k,i}^R\right)\left(\frac{1}{n}\sum_{i=1}^{n}X_{j,i}^R\right)\right]\nonumber\\
  &=&\frac{1}{n}\sum_{i=1}^{n}X_{k,i}^RY_{k,i}^R-\left(\frac{1}{n}\sum_{i=1}^{n}X_{k,i}^R\right)\left(\frac{1}{n}\sum_{}^{}Y_i^R\right),\ \ k=1,\cdots,p.\label{eqn-8}
\end{eqnarray}
Writing equations (\ref{eqn-8}) in matrix form yields (\ref{solu-gamma}). (\ref{solu-theta}) is obtained by plugging $\gamma_j$, $j=1,\cdots,p$ in equation (\ref{eqn-6}). 
\end{proof}

\subsection{Proof of Corollary \ref{coro:model-bias}}
\begin{proof}
From Theorem \ref{thm:lse-positivity}, $\hat{\gamma}={S_1}/{S_{1,1}}<0$, where $S_1$ and $S_{1,1}$ are the sample covariance of $X^R$ and $Y^R$, and the sample variance of $X^R$, respectively. Namely,
\begin{eqnarray}
  &&S_{1}=\frac{1}{n}\sum_{i=1}^{n}\left(Y_i^R-\overline{Y^R}\right)\left(X_i^R-\overline{X^R}\right)<0,\label{coro6:eqn-1}\\
  &&S_{1,1}=\frac{1}{n}\sum_{i=1}^{n}\left(X_i^R-\overline{X^R}\right)^2>0.\label{coro6:eqn-2}
\end{eqnarray}
Let $\left[\tilde{\gamma},\tilde{\theta}\right]$ be the joint constrained LS estimates of $\left[\gamma, \theta\right]$ such that $\tilde{\gamma}\geq 0$. Then, 
\begin{equation*}
  \tilde{\theta}=\overline{Y^R}-\tilde{\gamma}\overline{X^R}. 
\end{equation*}
It follows that
\begin{equation*}
  \tilde{Y}_i^R=\tilde{\gamma}X_i^R+\tilde{\theta}=\tilde{\gamma}X_i^R+\left(\overline{Y^R}-\tilde{\gamma}\overline{X^R}\right)
  =\overline{Y^R}+\tilde{\gamma}\left(X_i^R-\overline{X^R}\right),\ \ i=1,\cdots,n.
\end{equation*}
Then the sum of squared errors for the prediction of $Y^R$ based on the constrained LS estimates is calculated to be
\begin{eqnarray*}
  \sum_{i=1}^{n}\left(Y_i^R-\tilde{Y}_i^R\right)^2
  &=&\sum_{i=1}^{n}\left[\left(Y_i^R-\overline{Y^R}\right)-\tilde{\gamma}\left(X_i^R-\overline{Y^R}\right)\right]^2\\
  &=&\sum_{i=1}^{n}\left(Y_i^R-\overline{Y^R}\right)^2+\tilde{\gamma}^2\sum_{i=1}^{n}\left(X_i^R-\overline{X^R}\right)^2
  -2\tilde{\gamma}\sum_{i=1}^{n}\left(Y_i^R-\overline{Y^R}\right)\left(X_i^R-\overline{X^R}\right)\\
  &=&\sum_{i=1}^{n}\left(Y_i^R-\overline{Y^R}\right)^2+\tilde{\gamma}^2nS_{1,1}-2\tilde{\gamma}nS_1.
\end{eqnarray*}
Therefore, in view of (\ref{coro6:eqn-1})-(\ref{coro6:eqn-2}), 
\begin{equation}
  \sum_{i=1}^{n}\left(Y_i^R-\tilde{Y}_i^R\right)^2\geq\sum_{i=1}^{n}\left(Y_i^R-\overline{Y^R}\right)^2,
\end{equation}
and ``='' holds if and only if $\tilde{\gamma}=0$. This completes the proof. 
\end{proof}

\subsection{Proof of Corollary \ref{coro:lse-positivity}}
\begin{proof}
Under Assumption \ref{assumption-1},
\begin{equation*}
  \Sigma_{\textbf{X}^R}\to diag\left\{\text{Var}\left(X_1^R\right),\cdots,\text{Var}\left(X_p^R\right)\right\}\ a.s.,
\end{equation*}
and
\begin{equation}\label{coro5:eqn-1}
  \Sigma_{\textbf{X}^R,\textbf{Y}^R}\to
	\left[\text{Cov}\left(X_1^R,Y^R\right),\cdots,\text{Cov}\left(X_p^R,Y^R\right)\right]^T\ a.s..
\end{equation}
It follows that 
\begin{equation}\label{coro5:eqn-2}
  \hat{\Gamma}\to
	\left[\frac{\text{Var}\left(X_1^R\right)}{\text{Cov}\left(X_1^R,Y^R\right)},\cdots,
	\frac{\text{Var}\left(X_p^R\right)}{\text{Cov}\left(X_p^R,Y^R\right)}\right]^T\ a.s..
\end{equation}
Equations (\ref{coro5:eqn-1}) and (\ref{coro5:eqn-2}) together imply
\begin{equation*}
  \hat{\gamma}_jS_j\to\text{Var}\left(X_j^R\right)\ \ a.s.,\ j=1,\cdots,p,
\end{equation*}
and therefore,
\begin{equation*}
  P\left(\hat{\gamma}_jS_j>0\right)\to 1,\ \ \text{as}\ n\to\infty,\ j=1,\cdots,p.
\end{equation*}
Hence, if $S_j>0$, 
\begin{equation*}
  P\left(\hat{\gamma}_j>0\right)\to 1, \ \ \text{as}\ n\to\infty. 
\end{equation*}
\end{proof}

\subsection{Proof of Theorem \ref{thm:r-positive}}
\begin{proof}
Notice that 
\begin{eqnarray*}
  E\left[\left(Y_i^R-\hat{Y}_i^R\right)\hat{Y}_i^R\right]
  &=&E\left\{E\left[\left(Y_i^R-\hat{Y}_i^R\right)\hat{Y}_i^R|X_i^R\right]\right\}\\
  &=&E\left[\hat{Y}_i^RE\left(Y_i^R-\hat{Y}_i^R|X_i^R\right)\right]\\
  &=&0.
\end{eqnarray*}
Therefore,
\begin{equation*}
  E\left(Y_i^R\right)^2=E\left(Y_i^R-\hat{Y}_i^R\right)^2+E\left(\hat{Y}_i^R\right)^2.
\end{equation*}
This together with the fact that $E\left(\hat{Y}_i^R\right)=E\left(Y_i^R\right)$ yields
\begin{equation}\label{thm3:eqn-1}
  E\left(Y_i^R-\hat{Y}_i^R\right)^2=E\left(Y_i^R\right)^2-E\left(\hat{Y}_i^R\right)^2
  =\text{Var}\left(Y_i^r\right)-\text{Var}\left(\hat{Y}_i^R\right).
\end{equation}
Separately, 
\begin{equation}\label{thm3:eqn-2}
  E\left(Y_i^R-\hat{Y}_i^R\right)^2=E\left(\epsilon_i^R\right)^2=E\left(\epsilon_i^U-\epsilon_i^L\right)^2=2\sigma^2.
\end{equation}
By Markov's inequality, we have
\begin{equation}\label{thm3:eqn-3}
  P\left(\hat{Y}_i^R<0\right)\leq P\left(|\hat{Y}_i^R-Y_i^R|>Y_i^R\right)
  \leq\frac{E\left(Y_i^R-\hat{Y}_i^R\right)^2}{\left(Y_i^R\right)^2}.
\end{equation}
(\ref{thm3:eqn-3}) together with (\ref{thm3:eqn-1}) and (\ref{thm3:eqn-2}) proves the desired result. 
\end{proof}




\end{document}